\documentclass[preprint,12pt]{elsarticle}

\usepackage{lineno}
\usepackage[english]{babel}
\usepackage[utf8]{inputenc}
\usepackage{multirow}
\usepackage{amsmath}
\usepackage{hhline}
\usepackage{graphicx}
\usepackage{subcaption}
\usepackage{float}
\usepackage{hyphenat}
\usepackage[inline]{enumitem}
\usepackage[x11names]{xcolor}
\usepackage{listings}
\usepackage[ruled,vlined,linesnumbered]{algorithm2e}
\usepackage{tabularx,ragged2e}
\newcolumntype{C}{>{\Centering\arraybackslash}X}

\usepackage{tasks}
\usepackage{amssymb}
\newcommand{\MyPara}[1]{\vspace{.2em}\noindent\textit{\textbf{#1}}\hspace{.3em}}
  {\begin{list}{}{\setlength{\labelwidth}{0pt}
   \setlength{\itemindent}{0pt}
   \setlength{\listparindent}{\parindent}
   }}%
  {\end{list}}

\journal{Knowledge-Based Systems}

\begin{document}

\begin{frontmatter}


\title{Building Energy Consumption Models Based On Smartphone User's Usage Patterns}



\author[ufpe,ifpe]{Ant\^{o}nio S\'{a} Barreto Neto\corref{cor1}}
\ead{acsabarreto@gmail.com}
\ead[url]{http://www.modcs.org}

\author[ifpe]{Felipe Farias}
\ead{felipefariax@gmail.com}

\author[ufpe]{Marco Aur\'{e}lio Tomaz Mialaret}
\ead{marcomialaret@gmail.com}
\ead[url]{http://www.modcs.org}

\author[ufpe,ifpe]{Bruno Cartaxo}
\ead{email@brunocartaxo.com}
\ead[url]{http://www.brunocartaxo.com}

\author[ufpe]{Pr\'{i}scila Alves Lima}
\ead{pal3@cin.ufpe.br}
\ead[url]{http://www.modcs.org}

\author[ufpe]{Paulo Maciel\corref{cor2}}
\ead{prmm@cin.ufpe.br}
\ead[url]{http://www.modcs.org}

\cortext[cor1]{Principal Corresponding author}
\address[ufpe]{Federal University of Pernambuco (UFPE), Av. Jornalista Aníbal Fernandes, s/n - Cidade Universitária, Recife - PE}
\address[ifpe]{Federal Institute of Pernambuco - Paulista, Pernambuco, Brazil}

\begin{abstract}
The increasing usage of smartphones in everyday tasks has been motivated many studies on energy consumption characterization aiming to improve smartphone devices' effectiveness and increase user usage time. In this scenario, it is essential to study mechanisms capable of characterizing user usage patterns, so smartphones' components can be adapted to promote the best user experience with lower energy consumption.
The goal of this study is to build an energy consumption model based on user usage patterns aiming to provide the best accurate model to be used by application developers and automated optimization.
To develop the energy consumption models, we established a method to identify the components with the most influence in the smartphone's energy consumption and identify the states of each influential device. Besides that, we established a method to prove the robustness of the models constructed using inaccurate hardware and a strategy to assess the accuracy of the model built.
After training and testing each strategy to model the energy consumption based on the user's usage and perform the Nemenyi test, we demonstrated that it is possible to get a Mean Absolute Error of 158.57mW when the smartphone's average power is 1970.1mW.
Some studies show that the leading smartphone's workload is the user. Based on this fact, we developed an automatic model building methodology that is capable of analyzing the user's usage data and build smart models that can estimate the smartphone's energy consumption based on the user's usage pattern. With the automatic model building methodology, we can adopt strategies to minimize the usage of components that drain the battery.
\end{abstract}

\begin{keyword}
Android Service, Deep Neural Networks, Energy Consumption, Machine Learning, Neural Networks, Recurrent Neural Networks, User usage pattern.

\end{keyword}

\end{frontmatter}


\section{Introduction}
\label{sec:introduction}

It is not news that smartphones are now one of the leading personal devices, being used by billions of people, and even exceeding desktop computers \cite{smart:2017:Online}.In 2019 the StatCounter published statistics about the operating systems' usage and showed that Android is the leading with 39.19\% of market share \cite{statCounter:2019:Online}. This trend, however, brings its challenges. One of the most prominent is the smartphones high energy consumption, which is worsening due to our current limited battery lifespan. In 2005, Paradiso \textit{et al.}\cite{paradiso2005energy} developed a study that demonstrated that the battery technology evolution is not at the same pace as the other smartphone's devices.

A study carried out by Phone Arena \cite{phoneArena:2017:Online} obtained the historical data of the battery life over three years. This research showed that the evolution of the average battery lifespan varied considerably. This fact emphasizes the significant energy demand by the smartphone components and the ongoing necessity for manufacturers to develop techniques that improve the power consumption efficiency.

In 2009, Shye \textit{et al.} \cite{shye2009into} showed that the user is the primary workload contributing to smartphone usage; thus, it is essential to consider the user behavior when optimizing the power consumption of mobile phones. 
Wang \textit{et al.}\cite{wang2013power}, in 2013, proposed a profile-based battery traces methodology. According to them, battery traces can be easily collected through a user-level application on any device. Although it is difficult to achieve accurate results for only a few users because battery changes are coarse-grained, the method is expected to reach an accurate estimation when the number of battery traces reaches a certain scale. 
In 2015, Kim \textit{et al.} \cite{kim2015smartphone} proposed a framework to carefully monitor system events initiated by user interactions and identify the current user activity based on an online activity model.

Besides the energy consumption modeling, many studies in the research area of activity prediction modeling have been made to evaluate the context in which the users use their smartphones. 

Bejani \textit{et al.}\cite{bejani2018context} created a system to evaluate the driving style using the smartphone's sensors. Gallego \textit{et al.}\cite{gallego2013evaluating} use Android smartphones to purpose a context-aware restaurant recommender. Their developed software evaluates the impact of proactivity in the user experience.

Based on the findings of the previous studies, which demonstrate the importance of studying strategies to extend the smartphone's usage time due to its applicability in many situations, we are motivated to investigate approaches to extend the smartphone's usage time. Besides that, studying the user behavior to model the smartphone's energy consumption to improve it, we undertook a study to build an Energy Consumption Model based on the user's usage pattern. Our research aims to provide the most accurate model to be used by application developers and automated optimization. The developed models will provide the most accurate predictions than the Android battery historian that considers only the current user's usage.

To develop energy consumption models, we raise the following questions:
\fbox{\parbox{\dimexpr\linewidth-2\fboxsep-2\fboxrule\relax}{\textbf{RQ1:} Is the user the primary workload for mobile systems?\\\textbf{RQ2:} What the components with the most influence in the smartphone's energy consumption?\\ \textbf{RQ3:} How well does our model generalize to any device?\\ \textbf{RQ4:} How to assess the accuracy of the model built?}}
After posing these questions, we established a method to answer each one of them.

During this research, we demonstrated that the user's usage influences in smartphone energy consumption so much, as discussed in Section \ref{sec:userInfluenceEnergyConsumptionMethod}. Besides that, we found the smartphone's devices, which most influence energy consumption, as demonstrated in Section \ref{sec:dataAnalysisEvaluation}. In Section \ref{sec:energyConumptionModelGeneralizationEvaluation} we demonstrate how we can generalize our energy consumption models and evaluate them in Section \ref{sec:energyConsumptionModelsEvaluation}. The energy consumption models developed during this research demonstrated that it is possible to get a Mean Absolute Error of 158.57mW when the smartphone's average power is 1970.1mW. Besides that, we demonstrated during this research that, even in the presence of faulty hardware, we get a Mean Absolute Error of 268.42mW when the smartphone's average power is 1523.62mW.
    
In summary, the main \textbf{contributions} of this research are:
    
\begin{itemize}
    
    \item Study the user's usage pattern through an Android service which is capable of logging the smartphone's devices' states;
    \item Study the smartphone energy consumption based on the user usage pattern;
    \item Build an Energy Consumption model that uses the information provided by the log and the smart models to estimates the smartphone's energy consumption.
    \item Purpose a modeling methodology capable of providing the smartphone's devices energy consumption based on the user's usage pattern.
    \item Develop an Android application to record device states to be used as input to the energy consumption model based on the user's usage developed during this research.
    \item Develop and deploy an automated methodology capable of receiving the user's usage data, analyzes the received data,  and prepares the user's usage data to be used by a smart model. Besides that, the developed automated methodology chooses the most appropriate model to estimates the smartphone's energy consumption based on the user's usage pattern.
\end{itemize}

The remainder of this paper is organized as follows. Section  \ref{sec:relatedWorks} presents the related works.  Section \ref{sec:background} shows the background used as a basis for the development of this study. Section \ref{sec:proposedMethod} shows the proposed method to study the user's usage characteristics that should be considered to build an automatic model building mechanism that creates the estimators of the smartphone's energy consumption. In Section \ref{sec:automaticModelMethod}, we show the methodology used to, based on the studies, create the deployable method to build automatically the models that will be used to estimates the smartphone's power consumption based on the user's usage pattern.   Section \ref{sec:studyMethodDeployment} provides the deployment of the infrastructure to study the user's usage characteristics that should be considered to build an automatic model building mechanism that creates the estimators of the smartphone's energy consumption. Section \ref{sec:automaticModelEvaluation} discuss the strategies used to deploy the necessary modules to create automatically the models that estimate the smartphone's energy consumption. In
Section \ref{sec:discussion} we discuss about the research questions listed above. Finally,  Section \ref{sec:conclusion} presents the conclusion and future works.

\section{Related Works}
\label{sec:relatedWorks}

Many researchers conduct studies aiming to model the smartphone`s energy consumption using software engineering techniques.

The strategy proposed by Romansky \textit{et al .}\cite{romansky2017deep} models the smartphone energy consumption using the system calls inside the application code as inputs to a Long Short Term Memory Neural Network. The authors used 6 Android Apps in their different versions to collect the system calls and provide them to the Neural Network to predict smartphone energy consumption. The authors have proven the effectiveness of their method, getting an error of only 4\% between the measured energy consumption and their model prediction.

Another approach used by the authors of the research area of the smartphone's energy consumption model is to develop profiler apps that instrumentalize the studied Android apps source code to exercise them and collect battery usage statistics data \cite{di2017software}\cite{hao2013estimating}\cite{pathak2011fine}. 

Using another approach, Hindle \textit{et al .}\cite{hindle2014greenminer} has proposed the Green Miner: dedicated testbed hardware that mining software repositories. The GreenMiner physically measures the energy consumption of mobile devices (Android phones) and automates the testing of applications and the reporting of measurements back to developers and researchers.

Another strategy used to build a smartphone's energy consumption model was proposed by Alawnah and Sagahyroon\cite{alawnah2017modeling}, which offered an energy consumption model using the user's usage as inputs to a Multilayer Perceptron Network. The authors obtained a Root Mean Squared Error in order of $0.21$. During their work, they related the usage of high precision hardware to get the instantaneous power because their model is subject to fluctuations in its precision due to hardware failure.

Besides energy consumption modeling, some strategies are used for mobile phone energy consumption optimization using usage pattern models. Sometimes these models have been developed to identify the smartphone user usage pattern to serve as a subsidy for advertising campaigns \cite{alfawareh2014smartphones} \cite{li2015characterizing}. Those studies are usually conducted to discover the \textit{User-Battery Interaction}. The term was first coined by Benerjee \textit{et al.} \cite{banerjee2007users}. They conducted a systematic user study on battery usage and recharged behavior on both laptop computers and mobile phones. Those are the most important findings:
\begin{enumerate*}
  \item batteries are mostly recharged when the battery level is substantial;
  \item a considerable portion of recharges are context-driven (in terms of location and time), and those driven by battery levels usually occur when the battery level is high; and 
\item there is significant variation among users and systems.
\end{enumerate*}
To reach those results, they designed, deployed, and evaluated a user- and statistics-driven energy management system, named Llama, to exploit the battery power in a user-adaptive and user-friendly fashion to more optimally serve the user. The methodology is based on three strategies:  
\begin{enumerate*}
\item a passive logging tool that periodically records the battery level and charging status;
\item interviewing users to obtain qualitative data regarding users' battery usage, and
\item determining \textit{in situ} users' motivation for recharging their system.
\end{enumerate*}

According to Tarkoma \textit{et al.} \cite{tarkoma2014smartphone}, many studies on charging behavior have been conducted to date. All of them agree on a few salient points. First, recharging is either triggered by the current battery level or by the context, which is derived from the time and/or location and leads to two distinct user types.

Tarkoma \textit{et al.} \cite{tarkoma2014smartphone} also discussed the extent to which battery awareness applications change user behavior citing the example of Athukorala \textit{et al.} \cite{athukorala2014carat}, who surveyed 1,140 Carat users by asking them how the app changed their behavior. The results showed that users who used the app for more than three months were much more likely to stop using applications identified as high consumers by Carat.

Vallina-Rodriguez \textit{et al.} \cite{vallina2010exhausting} discussed the interdependence caused by the applications and users' behavior and their effect on battery life. The authors' results indicate that simple algorithmic and rule-based scheduling techniques are not the most appropriate for managing the resources because their usage can be affected by contextual factors. 

Another strategy proposed by the researchers in this area uses machine learning and neural networks to identify the context and predict the user activity using classifiers\cite{xie2017recognizing},\cite{dai2015sequential},\cite{bettini2020caviar}.

Xie \textit{et al.}\cite{xie2017recognizing} use machine learning strategies to identify the context of an e-learning platform usage to adapt the content served to the learners. The developed tool uses sensors like accelerometer, light, and sound to determine the usage context.

Dai \textit{et al.}\cite{dai2015sequential} uses machine learning algorithms to predicts the activity performed by a user. To aim this objective, the authors use the cross-user activity transfer technique that uses similar activities performed by other users to augment the machine learning algorithm capacity of predicts the activity performed by a user.

Bettini \textit{et al.}\cite{bettini2020caviar} built a system, CAVIAR(Context-aware ActVe and Incremental Activity Recognition), which combines semi-supervised learning and semantic context-aware reasoning. The authors purpose the system to predicts the user activity using an incremental strategy, in the same way, that we approach the problem of constructing the smartphone's energy consumption in this research.

Prior work has created a smartphone's energy consumption model based on users' usage pattern\cite{alawnah2017modeling} similar to our goal. Furthermore, some energy models have been generated using dedicated hardware\cite{hindle2014greenminer}. Other prior works have been assumed that it is possible to interact with the Android App source code to identify the system calls with a significant impact on the smartphone's energy consumption\cite{romansky2017deep}.

However, our work does not assume that the user has access to the Android app source code or that the hardware responsible for measuring the instantaneous power is confident. During this research, we have constructed an automatic model building methodology capable of automatically analyze the user's usage data and build an energy consumption model capable of estimates the smartphone's energy consumption model based on the user's usage pattern. The automatic methodology developed in this work can be used by autonomous energy optimization platforms.

\section{Background}
\label{sec:background}

During this research, we undertook a study to build a Power Model based on the user's usage pattern aiming to optimize battery life span and minimize the usage of components that drain the battery.
To accomplish this objective, we need to define a strategy to capture the states of a device, log them, analyze the results, and determine a mechanism that optimizes the CPU usage, as explained in the related section. Here, we present the concepts necessary to understand the method we are proposing in this research.

\subsection{Android Core Concepts}
\label{sec:androidCoreConcepts}

Here we present the main concepts about the Android platform necessary to understand the method we propose in this research.

\MyPara{Services:} According to Android developers \cite{androidService:2020:Online}, a service is an application component that can perform long-running operations and does not provide a user interface. There are two ways to start a service: through an application component that can begin it or through an element that can bind it to interact with it and even perform interprocess communication (IPC).

\MyPara{Broadcasts:} Android apps can send or receive broadcast messages from the Android system and other Android apps \cite{androidbroadcast:2020:Online}. These broadcasts are sent when an event of interest occurs. Apps can register to receive specific broadcasts, and when a broadcast is sent, the system automatically routes the messages to apps subscribed to receive that particular type of broadcast \cite{androidbroadcast:2020:Online}.

\subsection{Procfs}
\label{sec:procfs}

The process file system (procfs) provided by the Linux
kernel offers process-specific run-time information \cite{linuxprogrammers:2019:Online}. For
example, the process utime feature records how long a process
has been scheduled in user land in clock ticks. The procfs
provides additional resource-usage features associated with a
running application that we can use to predict the energy
consumption of a given application.

\subsection{Nemenyi Test}
\label{sec:nemenyiTest}

According to Japkowicz and Shah\cite{japkowicz2011evaluating}, the Nemenyi test computes the \textit{q} statistic over the difference in average mean ranks of the classifier or regressor.

So, the necessary steps to perform the Nemenyi test are:
\begin{enumerate}
    \item Build a ranking for each data set using the performance measure in ascending order from the best performance algorithm to the worst performance algorithm;
    \item Let $R_{ij}$ be the rank of the algorithm $f_{j}$ on the data set $S_{i}$;
    \item We compute the mean rank of algorithm $f_{j}$ on all data sets using the expression $\overline{R_{j}}=\frac{1}{n}\sum_{i=1}^{n}R_{ij}$;
    \item For any two classifiers $f_{j1}$ and $f_{j2}$ we compute the \textit{q} statistic using the expression $q=\frac{\overline{R_{j1}} - \overline{R_{j2}}}{\sqrt{\frac{k(k+1)}{6n}}}$ in which \textit{k} is the number of evaluated algorithms and \textit{n} is the number of data sets;
    \item The null hypothesis is rejected after a comparison of the obtained \textit{q} value for the desired significance table for critical $q_{\alpha}$ values, where $\alpha$ refers to the significance level. Reject the null hypothesis if the obtained \textit{q} value exceeds $q_{\alpha}$.  
 \end{enumerate}

\subsection{Machine Learning and Neural Networks}
\label{sec:MLandNeuralNet}

To define the machine learning term, we used the one given by Mitchel\cite{mitchell1997machine}:
A computer program is said to learn from experience E with respect to some task T and some performance measure P, if its performance on T, as measured by P, improves with experience E.

In this way, during our research, we have adopted the workflow defined by  G\'{e}ron\cite{geron2019hands}. The author establishes a procedure to build the workflows when we are using machine learning to solve our problem defined by seven steps:

\MyPara{1. Look the Big Picture}: In this step, we have to evaluate the problem to choose the information that we have to extract to solve the problem. Besides that, we have to select the metrics that we will use to assess the performance of the chosen algorithms.

\MyPara{2. Get the data:} After defining the problem, we have to establish strategies to get the information that we selected in the previous step.

\MyPara{3. Discover and visualize the data to gain insights:} In this step, we have to visualize the data using plots, correlation matrices, and experimenting with attributes combinations.

\MyPara{4. Prepare the data for machine learning algorithms:} According to the author, we should build functions that process the data in a way that the operations used to transform the data should be reproducible. The primary activities that should be done in this step were: data clean, manipulation of textual and categorical attributes, and attribute scaling.

\MyPara{5. Select a model and train it:} According to the author, we should choose the most appropriate algorithms for our problem once we have information concerning the problem, and the data already been transformed for a group of algorithms.

\MyPara{6. Fine-tune your model:} In this step, after getting the preliminary results, the designer should adjust the parameters of the chosen algorithms to get the maximum performance of each one.

\MyPara{7. Present your solution:} It is the final step of the project in which we should present the results to make it feasible to deploy the solution.

\subsection{Mutual Information}
\label{sec:mutualInformation}

According to Brownlee\cite{brownlee2019probability}, Mutual Information is calculated between two variables and measures the reduction in uncertainty for one variable given a known value of the other variable.

The mutual information between two variables can be stated formally through the relation shown in Equation \ref{eqn:mutualInformation}. 

\begin{equation}
    I(X,Y) = H(X) - H(X|Y)
    \label{eqn:mutualInformation}
\end{equation}

Where $I(X,Y)$ is the mutual information for $X$ and $Y$, $H(X)$ is the entropy for $X$ and $H(X|Y)$ is the conditional entropy for $X$ given $Y$. This measure is symmetrical, meaning that $I(X,Y)=I(Y,X)$.

\section{Proposed Method}
\label{sec:proposedMethod}

\begin{figure}[ht]
	\centering
	\includegraphics[width=\textwidth]{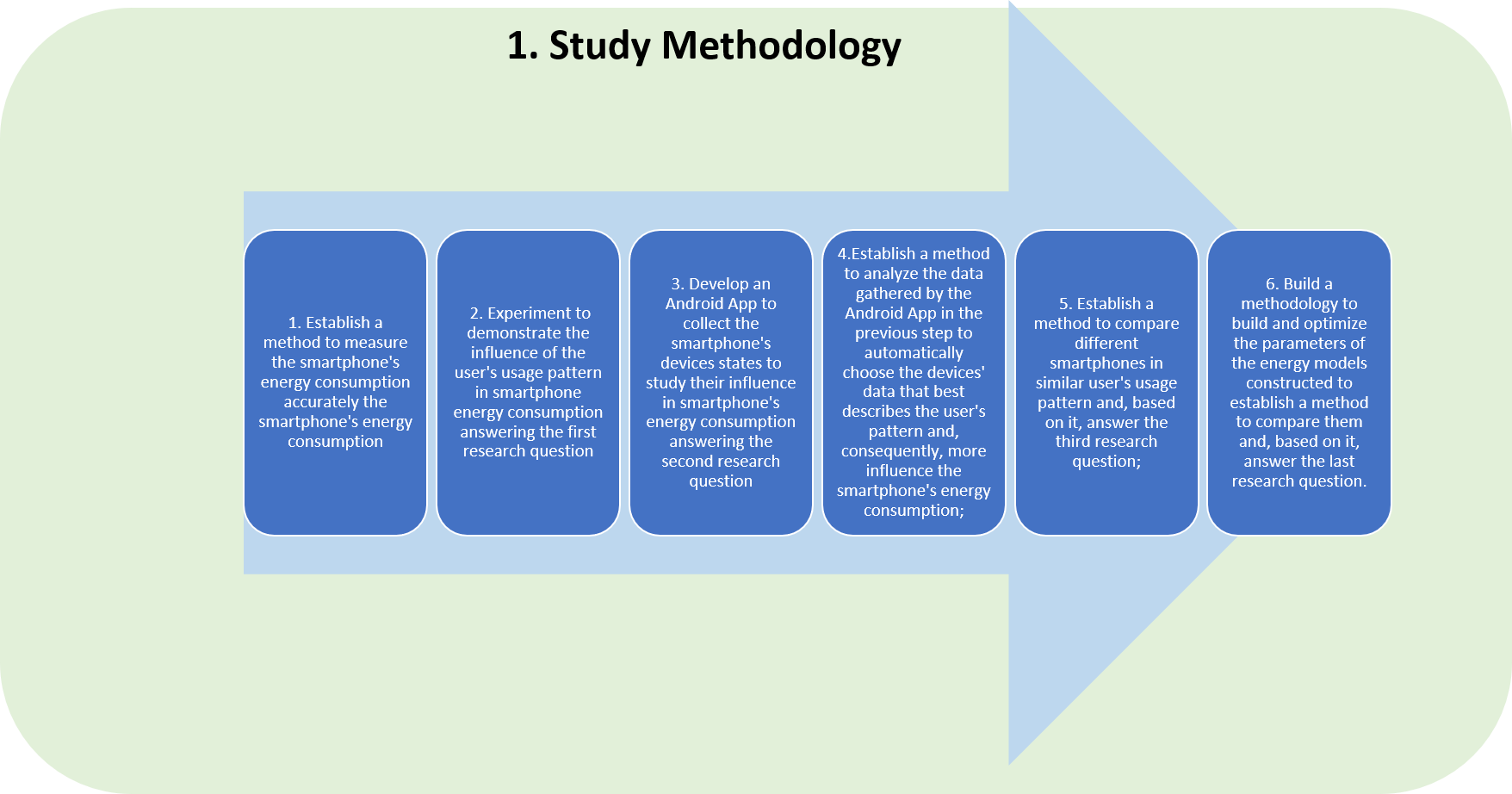}
	\caption{Study methodology to define the automatic model building methodology}
	\label{fg:studyMethodology}
\end{figure}

To define the methodology to automatically select the best model that estimates the smartphone's consumption based on the user's usage, we need to establish a method to study the relevant characteristics to define a consumption model based on the user's use. Besides that, we need to develop another procedure that we can implement to build the energy consumption model automatically.
The steps of the study methodology used to define the method capable of automatically generate the energy models are depicted in Figure \ref{fg:studyMethodology}.

In the next sections, we will discuss each step shown in Figure \ref{fg:studyMethodology}.

\subsection{Step 1: Energy Consumption Measurement}
\label{sec:cunsmptionMeasurementMethod}

To build energy consumption models in real-time, we need to establish a strategy to measure smartphone's instantaneous energy consumption in real-time.

To aim this goal, we used the expression shown in Equation \ref{eqn:energyConsumption}.
\begin{equation}
    E(t)=\int_{0}^{t} P(t)\;dt
    \label{eqn:energyConsumption}
\end{equation}
In Equation \ref{eqn:energyConsumption} the \textbf{E(t)} represents the energy consumption in a time \textbf{t} and \textbf{P(t)} represents the instantaneous power in a time \textbf{t}. So, based on the Equation \ref{eqn:energyConsumption}, we need to find a way to get the instantaneous smartphone power. To reach this goal, we used the property expressed by equation \ref{eqn:InstPower} in which the \textbf{U(t)} represents the instantaneous voltage function in time, and \textbf{I(t)} shows the instantaneous current function in time.

\begin{equation}
    P(t)=U(t)\times I(t)
    \label{eqn:InstPower}
\end{equation}

Based on equation \ref{eqn:InstPower}, we need to find a way to get the instantaneous voltage and current to accomplish our goal that is finding a way to measure the smartphone's current energy consumption.

Although the Android operating system offers the Battery Manager API\cite{batteryManager:2020:Online}, we used the \textit{/sys/class/power\underline{\space}supply/battery} Linux procfs for a more accurate current measurement. So, we used the  \textit{batt\underline{\space}current\underline{\space}ua\underline{\space}now} virtual file to get the instantaneous current in micro ampere. To measure the instantaneous voltage we used the \textbf{EXTRA\underline{\space}VOLTAGE} property of the Android Battery Manager API\cite{batteryManager:2020:Online}. 

With the instantaneous current and voltage, we get the instant power using the equation \ref{eqn:InstPower} and, with the instant power, we calculated the current energy consumption using equation \ref{eqn:energyConsumption} based on the fact that our sample rate is 1 second.

\subsection{Step 2: User's Usage Pattern influence in the smartphone's energy consumption}
\label{sec:userInfluenceEnergyConsumptionMethod}

The main contribution of this research is the construction of the smartphone's energy consumption model, considering the user's usage pattern. Based on it, before to establish the necessary steps to aim the primary goal of this research, we need to study the smartphone's energy consumption in similar usage conditions, differing only particular user's usage conditions.

To study the smartphone's energy consumption is similar usage conditions, we used the \textbf{Samsung Galaxy A10} because during our tests using it and the \textbf{Motorola Moto G6}, we identified that the \textbf{Galaxy A10} has better hardware to measure the instantaneous current and voltage. So, we set the display brightness, removed the sim card, enabled the Wi-Fi opened only the Youtube, and played two different video clips in two separate tests.

\begin{figure}
	\centering
	\includegraphics[width=0.6\textwidth]{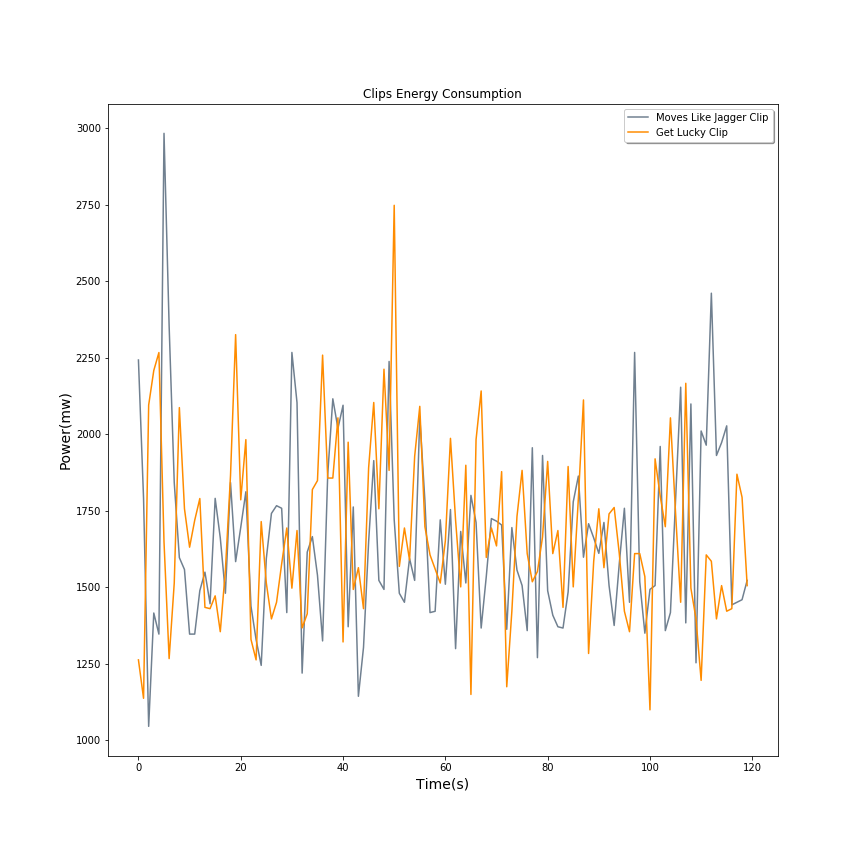}
	\caption{User's usage Experiment in Galaxy A10}
	\label{fg:clipsConsumption}
\end{figure}

After that, we measured the smartphone energy consumption using the methodology described in Section \ref{sec:cunsmptionMeasurementMethod} for each video clip and get the results shown in Figure \ref{fg:clipsConsumption}.

To better analyze the smartphone's energy consumption when playing each video clip, we framed the time series that represent it to the first two minutes.

Analyzing the graphs shown in Figure \ref{fg:clipsConsumption}, we can perceive that there are differences in smartphone's energy consumption. When we calculate the energy consumption when playing each video in the considered time frame, we get Moves Like Jagger clip consuming $198412.431$ mJ and Get Luck clip consuming $201147.921$ mJ. These facts justify considering the user's usage pattern when building a smartphone's energy consumption model.

\subsection{Step 3: Devices Monitor Service}
\label{sec:devicesMonitorServiceMethod}

Based on the experiment performed in the previous step, we demonstrated the influence of the user's usage over the smartphone's energy consumption.

So, we need to develop an Android service that is capable of profiling the user's usage to, based on the generated log, build an energy consumption model.

Before developing the profiler, we need to identify the components which influence in smartphone's energy consumption. So, we build the bash script shown in Listing \ref{lst:energyProfiler}  that logs the instant power and voltage using the Linux kernel module named power supply class \cite{powerSupplyClass:2019:Online}. We exercised the leading smartphone's devices and set up many of them while profiling energy consumption.

\begin{lstlisting}[language=bash, label={lst:energyProfiler},caption="Energy Consumption Profiler"]
!#/bin/bash

count=0
while [ $count -lt 120 ]
do
current=`cat /sys/class/power_supply/battery/current_now_ua`
voltage=`cat /sys/class/power_supply/battery/voltage_now`
echo "$current;$volatge" >> log.txt
sleep 1
count=$((count+1))
done
\end{lstlisting}

Based on the results of the performed experiments, we selected the following devices to monitor:

\begin{tasks}[label={\arabic*}](3)
\task CPU Frequency
\task Screen
\task Wi-Fi
\task Radio(3G/4G)
\task Bluetooth
\task Current App
\task Battery Charge
\task CPU Usage
\task CPU Temperature
\task Mobile Signal\\Strength
\task Wi-Fi Signal\\Strength
\task Mobile RX Bytes
\task Mobile TX Bytes
\task Wi-Fi RX Bytes
\task Wi-Fi TX Bytes
\task Kb Read Per Second (Disk)
\task Kb Write Per Second (Disk)
\task Kb Read(Disk)
\task Kb Write(Disk)
\task Swap In
\task Swap Out
\task Context Switches
\task Red Mean
\task Red Std
\task Green Mean
\task Green Std
\task Blue Mean
\task Blue Std
\task Brightness
\task Orientation
\end{tasks}

With the definition of the most influential devices in smartphone energy consumption, we created an Android Service that uses the appropriate APIs to collect the device's usage.
The Android intents \textbf{ACTION\underline{\space}SCREEN\underline{\space}OFF} and \textbf{ACTION\underline{\space}SCREEN\underline{\space}ON} provided the information about the \textbf{Screen} state and the Android intent \textbf{ACTION\underline{\space}BATTERY\underline{\space}CHANGED} provided the information about the smartphone \textbf{Battery charge}.

The Android WifiManager class provided the information about the \textbf{Wi-Fi} state and \textbf{Wi-Fi Signal Strength} while the Android BluetoothAdapter class provided the information about \textbf{Bluetooth} state. The Android ConnectionType class provided the information necessary to update the \textbf{Radio (3G/4G)} states.

To monitor the \textbf{Mobile Signal Strength}, we created a specialization of the Android PhoneStateListener class, which has the \textit{onSignalStrengthsChanged} method responsible for captures the changes in the phone signal strength.

The Android NetworkStat class was used to provide information about \textbf{Mobile RX Bytes}, \textbf{Mobile TX Bytes}, \textbf{Wi-Fi RX Bytes}, and \textbf{Wi-Fi TX Bytes}.
To get the screen orientation, we created a specialization of the Android OrientationEventListener class to capture the change in screen orientation using the \textit{onOrientationChanged} method.

The Generic Thermal SysFs driver\cite{thermalSysFs:2019:Online} provided the information about the \textbf{CPU Temperature} through the \textit{thermal\underline{\space}zone0/temp} virtual file.
To measure the \textbf{CPU Frequency} we used the CPU Performance Scaling\cite{CPUPerfKernel:2019:Online}. We have used the \textit{cpuinfo\underline{\space}cur\underline{\space}freq} virtual file to measure the current frequency of each CPU inside the smartphone.

The Iostat\cite{iostat:2019:Online} linux tool was used to get the information about\textbf{ Kb Read Per Second(Disk)}, \textbf{Kb Write Per Second(Disk)}, \textbf{Kb Read(Disk)}, and \textbf{Kb Write(Disk)}.
The Vmstat\cite{vmstat:2019:Online} linux tool was used to get the information about \textbf{Swap In}, \textbf{Swap Out}, and \textbf{Context Switches}.

The Linux Kernel Backlight\cite{backlight:2019:Online} system class is responsible for measure the brightness of the screen and provide it to our service.
To obtain the \textbf{CPU Usage} we used the CPUInfo class from AndroidCPU App\cite{androidCPU:2020:Online}.
To obtain the Mean and Standard Deviation of the pixel colors of the image shown by the display every second, we used opencv-android plugin\cite{androidOpenCV:2020:Online}.

To get the \textbf{Current App}, we used the Foreground App Checker for Android\cite{appchecker:2020:Online} library that is responsible for monitoring the foreground App in the most recent Android versions.
Besides the devices described above, we included in the profiler the current (ma) and voltage (v) using the procedure described in Section \ref{sec:cunsmptionMeasurementMethod}.

In Section \ref{sec:monitorServiceEvaluation}, we will discuss the operation of the created monitor service and the overhead in the smartphone's energy consumption caused by it.

\subsection{Step 4: Automatic Data Analysis}
\label{sec:automaticDataAnalysisMethod}

One of the goals of this research study, as discussed in Section \ref{sec:introduction}, is to develop an energy consumption model that can be used by autonomous optimization mechanisms.

So, to develop an energy consumption model that can be used by autonomous optimization mechanisms, it is essential to create a procedure in which the tools can use to analyze the user's usage data automatically. The developed method can be used to provide the appropriate information in a structured way to the model to get the most accurate results from it and will be detailed in Section \ref{sec:automaticModelMethod}.

\subsection{Step 5: Energy consumption model accuracy in biased data sets}
\label{sec:energyConumptionModelGeneralizationMethod}

To develop an Energy Consumption model, we need to establish a methodology that can be used in any device. So, we need to build a strategy to demonstrate the power of generalization of the method used to construct the model accurately.

According to the prior research studies that developed energy consumption models, we can perceive that one of the main threats to validation is the accuracy of the hardware responsible for measuring the smartphone's energy consumption.

Based on it, we have found a smartphone with an inaccurate Battery Fuel Gauge IC. The chosen equipment was the Motorola Moto G6(XT1925-3) with Android 8 that has a measurement circuit that updates the current at different frequencies at a time.

To assess the accuracy of the generated models, we need another smartphone with an accurate Battery Fuel Gauge IC. The chosen equipment was the Samsung Galaxy A10(A105M) with Android 9 that updates the current most frequently than our 1-second sample rate.

To demonstrate the quality of the generated models, we performed the following steps:

\begin{enumerate}
    \item Perform the same experiments using the Samsung Galaxy A10 and the Motorola Moto G6;
    \item Get the frequencies in which the Motorola Moto G6 Battery Fuel Gauge IC repeats the value of current;
    \item Generate another Samsung Galaxy A10 data set simulating the behavior found in the Motorola Moto G6 data set;
    \item Train the energy consumption models using the biased Galaxy A10 data set;
    \item Test the energy consumption models using another unbiased Galaxy A10 data set;
    \item Compare the obtained results with a persistent model that repeats the last measurement and demonstrates that, even in a biased data set, our energy consumption models have an excellent performance.
\end{enumerate}

In Section \ref{sec:energyConumptionModelGeneralizationEvaluation} we will evaluate the accuracy of the generated models over biased data sets.

\subsection{Step 6: Energy Consumption Model Construction}
\label{sec:energyConumptionModelConstructionMethod}

To build the smartphone's energy consumption models considering the user's usage pattern, we need to adopt strategies that can learn and adapt according to the usage context.

So, we adopted a design methodology to aim the objective of this research: build an energy consumption model based on the user's usage pattern. The steps of the method are described in Section \ref{sec:automaticModelMethod}.

Following the steps described in Section \ref{sec:automaticModelMethod}, we can automate the choice procedure of the most appropriate strategy used to provide the smartphone's energy consumption model according to the user's usage pattern.

\section{Automatic model building methodology}
\label{sec:automaticModelMethod}

\begin{figure}[ht]
	\centering
	\includegraphics[width=\textwidth]{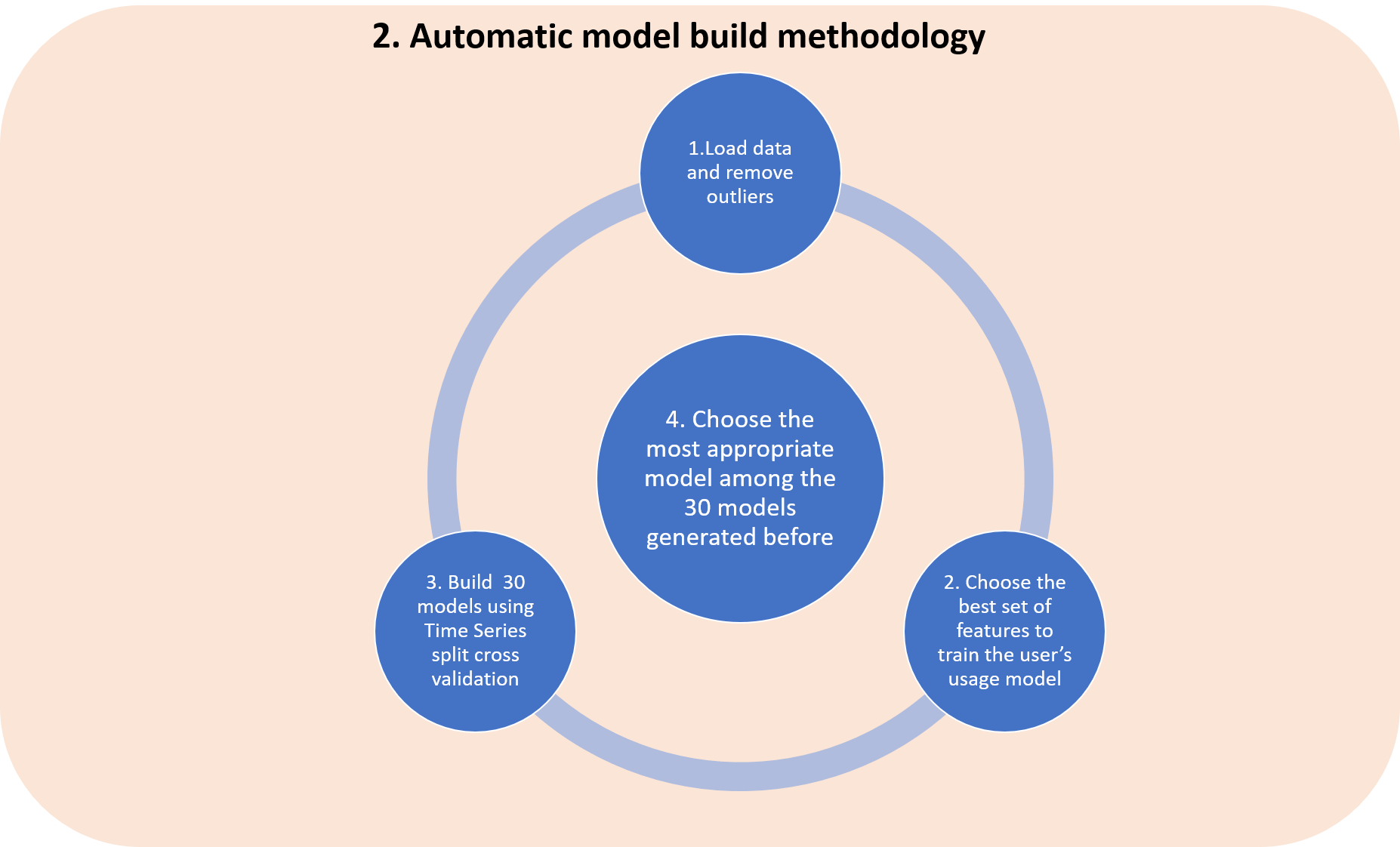}
	\caption{Methodology adopted to build the smartphone consumption models based on user's usage in an automatic way}
	\label{fg:modelBuildMethodlogy}
\end{figure}

In the previous section, we investigated the steps that we should follow to build a methodology which we capable of implementing to develop the smartphone's consumption model based on the user's usage automatically. Now we can establish the automatic model build methodology defining the steps depicted in Figure \ref{fg:modelBuildMethodlogy}.

In the next sections, we will discuss each step shown in Figure \ref{fg:modelBuildMethodlogy}.

\subsection{1. Load data and remove outliers}
\label{sec:loadDataMethod}
To develop the smartphone's energy consumption model based on the user's usage pattern, we need to load the user's usage data in a structured way analyzing it to prevent the presence of outliers. So, we need to develop a layer in our deployed solution responsible for reading the data transmitted by the user's smartphone in a structured way, run an automatic mechanism to analyze the provided data, and filter the outliers.

In this way, after performing some tests using the user's usage data, we developed a method to read the data provided by Devices Monitor Service following the procedure described below:
\begin{enumerate}
    \item Load the data uploaded by the devices monitor service;
    \item Analyze the features that have a single value to remove them;
    \item Analyze the features that have very few values to remove them;
    \item Analyze the features that have a low variance to remove them;
    \item Use the Local Outlier Factor, an automatic outlier detection mechanism, to remove the outliers from the user's usage data.
\end{enumerate}
Using this procedure, we structure the user's usage data to proceed to the next step in our automatic model build methodology. In the next step, we perform some actions to select the most representative features to build the smartphone's consumption model based on the user's usage pattern.

\subsection{2. Choose the best set of features to train the  user's usage model}

After applying some operations in the user's usage raw data in the previous step, we need to define a strategy to select a set of features that are more informative to the model estimates the smartphone's energy consumption based on the user's usage pattern.

To aim this objective, we defined a method described in Algorithm \ref{lst:bestKFeaturesAlgorithm}.
In the function \textbf{Select\_K} described in Algorithm \ref{lst:bestKFeaturesAlgorithm}, we iterate over the set of features performing a ten split cross-validation in each one to evaluates the estimation error.

In the function \textbf{build\_k\_rank}, we ranked the number of selected features based on the median errors produced by each one.
In lines 11 to 13, we performed 30 experiments evaluating cross-validation errors over the train data set.

In lines 15 to 19, we calculated the median errors of each set of selected attributes over the 30 experiments performed before.
On lines 20 to 24, we calculated the median, the 25th percentile, and the 75th percentile of the medians' values of the features set. Besides that, we figured the interquartile range(iiq).
After performing these calculations, on lines 26 to 28, we removed the features that have median errors bellow the low (median - 1.5* iiq) and above the high(median + 1.5*iiq) values.
So, based on these calculations, we returned the number of features that have the median error in the mean position of the filtered medians errors set.
The parameters used by the functions are:

\MyPara{1.X:}The features of the train data set that will be used as predictors by the estimator.
\\
\MyPara{2.y:}It represents the target feature of the train data set that will be used by the estimator.
\\
\MyPara{3.model:}The smart model that will be used to evaluate the set of predictor features of the train data set that minimizes the estimation errors.
\\
\MyPara{4.strategy:} The score function that will be used to select the features that most contribute to estimates the smartphone's power consumption.

\begin{algorithm}
\LinesNumbered
\SetAlgoVlined
\SetKwProg{Fn}{Function}{:}{}
\SetKwFunction{FselectK}{Select\_K}
\SetKwFunction{FbuildKRank}{build\_k\_rank}
\Fn{\FselectK{X,y,model,strategy}}{
$num\_features$=\FuncSty{columns(\ArgSty{X})}\\
$results=$\FuncSty{list()}\\
\For{$i=1$;$i\leq num\_features;i++$}{
$ppl=\FuncSty{Pipeline}([scaler,feature\_selector(\ArgSty{i},\ArgSty{strategy}),model])$
$scores = \FuncSty{coss\_validation}(estimator=ppl, splits=10,X,y)$\\
$results.add(\FuncSty{tuple}(i,scores))$
}
\KwRet{$results$}
}
\Fn{\FbuildKRank{X,y,model,strategy}}{
    $scores$=\FuncSty{Map<$key$,$value$>}\\
    $num\_features$=\FuncSty{columns(X)}\\
    \For{$i=0$;$i< 30;i++$}{
        results = \FuncSty{Select\_K}(X,y,model)\\
        scores[i]=results
    }
    $medians=$\FuncSty{list()}\\
   \For{$i=1$; $i \leq num\_features$; $i++$}{
    $feature\_medians$=\FuncSty{list()}\\
    \For{$j=0$; $j< 30$; $j++$}{
       $score\_array$= $scores[j][i]$
       $feature\_medians$.add(\FuncSty{median}($score\_array[1]$))
    }
    $medians$.add(\FuncSty{tuple($i$,\FuncSty{median($feature\_medians$)})})
   }
   
   $med$ = \FuncSty{median($medians[:][1]$)}\\
   $q25$,$q75$ = \FuncSty{percentiles($medians[:][1]$,[$25$,$75$])}\\
   $iiq$ = $q75$-$q25$\\
   $low$=$med -1.5 \times iiq$\\
   $high$=$med +1.5 \times iiq$\\
   
   $filtered\_medians$=\FuncSty{list()}\\
   \ForEach{$m$ in $medians$}{
        \If{$m[1]> low$ and $ m[1] < high$}{
            $filtered\_medians$.add($m$)
        }
   }
   $pos$ =\FuncSty{len($filetered\_median$)}/2\\
   $best\_k$=$filtered\_medians[pos][0]$\\
   
   \KwRet{$best\_k$}
}
 
 \caption{Select best set of features}
\label{lst:bestKFeaturesAlgorithm}
\end{algorithm}

\subsection{3. Build 30 models using Time Series Split cross validation}
\label{sec:modelsRepetitionMethod}
After performing the analysis of the data to remove the outliers in the first step and find the value of the \textbf{k} to select the most representative data to estimates the power consumption in the second step, we can estimate the smartphone's energy consumption based on the user's usage data.

So, in this step, we train the models with and without feature selection 30 times using a thirty split Time Series cross-validation to evaluate the most appropriate model among them to use as the final estimator.

To evaluate and choose the most appropriate in each of the thirty runs, we followed the procedure described below:
\begin{enumerate}
    \item Run the cross-validation with time series split;
    \item Evaluate the 30 test scores gathered in the previous step calculating the median, 25th percentile, and 75th percentile;
    \item Establish the cut-off point using the interquartile range;
    \item Remove the test scores bellow the $median - 1.5 \times iiq$ and above the $median + 1.5 \times iiq$
    \item Get the train size with the most samples inside the filtered set gathered in the previous step;
    \item Fit the model with the train size gathered in the previous step. 
\end{enumerate}

\begin{figure}
	\centering
	\includegraphics[width=\textwidth]{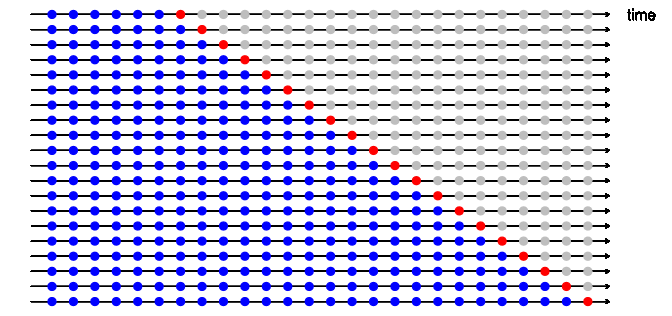}
	\caption{Time Series Cross Validation}
	\label{fg:timeSeriesCrossValidation}
\end{figure}

We have chosen using the Time Series cross-validation, as shown in Figure \ref{fg:timeSeriesCrossValidation} because the smartphone's energy consumption based on a user's usage pattern has a temporal correlation nature. It imposes an order in the train and test data that will be used to train and test the model during the cross-validation procedure. In Figure \ref{fg:timeSeriesCrossValidation} the train data set are split in different train (blue part) and test(gray part) parts over time. 

\subsection{4. Choose the most appropriate model}
\label{sec:modelsChoiceMethod}
In this stage, the automatic model build method evaluates various models using various configurations and features selection methods to assess these generated models.

After evaluating a set of models with and without feature selection using the previously described methods, we can assess the most appropriate model to estimates the smartphone's energy consumption based on the user's usage pattern.

To aim this objective, we used the Nemenyi's test, as described in Section \ref{sec:nemenyiTest}, to rank the scores of the evaluated models in a test data set and chooses the one that has the minimum error.

\section{Proposed Method Deployment}
\label{sec:studyMethodDeployment}

With the research methodology created according to described in Sections \ref{sec:proposedMethod} and \ref{sec:automaticModelMethod}, we need to structure an infrastructure to realize the experiments and generate the research results to demonstrate how we can automate the process of modeling the smartphone's energy consumption.

In this section, we will discuss the infrastructure created to realize the experiments and the procedures adopted to implement the study methodology, as described in Section \ref{sec:proposedMethod}.

In Section \ref{sec:automaticModelEvaluation}, we will discuss how we deployed the automatic model building methodology, as mentioned in Section \ref{sec:automaticModelMethod}, to build the smartphone's energy consumption model based on the user's usage pattern.

\subsection{Hardware/Software Infrastructure}
\label{sec:hardwareEvaluation}

To implement the methodology described in Sections \ref{sec:proposedMethod} and \ref{sec:automaticModelMethod}, we need to create a hardware/software infrastructure that is capable of profiling the devices described in Section \ref{sec:devicesMonitorServiceMethod}. After it, the developed infrastructure should process devices' usage according to the user's usage pattern and provides the information necessary to create the energy consumption model.

To simulate the user's usage pattern in different devices and configuration sets, we chosen two smartphones: Motorola Moto G6(XT1925-3) and Samsung Galaxy A10(A105M).

According to Phone More\cite{motog6:2020:Online}, the Motorola Moto G6(XT1925-3) has 32GB of internal memory, 3GB of RAM Memory LPDDR3, and its operating system is Android 8.0 (Oreo). Despite the possibility of upgrading its firmware, we decided to do not to alter it because we want to test our solution in different operating systems too.

Another smartphone used during our simulation was the Samsung Galaxy A10(A105-M). According to Phone More\cite{galaxyA10:2020:Online}, the Samsung Galaxy A10 has 32GB of internal memory, 2GB LPDDR4X, and its operating system is Android 9.0(Pie).

To process the user's usage data, we decided to use a desktop computer with a dedicated Nvidia GPU. So, we transformed the data generated during our experiments using a desktop computer with the following hardware: Intel Core I3-8100 CPU, Asus TUF H310M-Plus Gaming, 24 GB Ram Memory DDR4 2400 MHZ, and an NVIDIA RTX 2060.

To receive the log file containing the user's usage data, we used the Flask\cite{flask:2020:Online} lightweight WSGI web application framework for python with the Flask RESTful\cite{flaskRestful:2020:Online} plugin. It provides an API to receive the user's usage data, processes it, generates the energy consumption model, and delivers the model back to the smartphone.

To Analyze the data received, we used the Pandas Data Analysis Library \cite{pandas:2019:Online}. We used the DataFrame object that represents a structured data set, and, with it, we can convert columns and calculate the power in \textit{mw} using the current and the voltage logged. Besides that, we used our developed library to remove the outliers using the LocalOutlierFactor\cite{localOutlierFactor:2020:Online}, and select the \textbf{k} best features.

After analyzing the data, we have subsidies to generate energy consumption models. To aim with this objective, we used two approaches: machine learning and neural networks.

To build the machine learning models and perform feature selection, we used the Scikit-learn\cite{scikit-learn}, which is a well-designed machine learning library for python.

Despite the scikit-learn also supports neural networks implementing a  Multilayer Perceptron architecture, we used the Tensorflow\cite{tensorflow2015-whitepaper} because we tested not only a Multilayer Perceptron architecture but also a Recurrent Neural Network architecture.

\subsection{Devices Monitor Service Evaluation}
\label{sec:monitorServiceEvaluation}

To implement the methodology described in Section \ref{sec:devicesMonitorServiceMethod} using the APIs related before, we created an Android App named Devices Monitor Service.

The Devices Monitor Service app has monitors on the following devices: Bluetooth, CPU, Radio (3G/4G), Screen, and WiFi. Besides that, we have created two other monitors: App monitor and State monitor.
The App monitor is responsible for monitoring the foreground Android app.

The State monitor is responsible for summarizing the information from all the other monitors. Besides that, it is responsible for collecting the information about the all other devices described in Section \ref{sec:devicesMonitorServiceMethod} that do not have a specific monitor implemented, and also, the State monitor collects the instant current in \textit{ma} and voltage in volts.

To collect the information about the other devices which do not have a monitor implemented inside the Android app, the State monitor interacts with Linux tools through a command-line library named Android Shell\cite{androidShell:2020:Online}.

The data collected by the State monitor is saved in a CSV file to transmit it at the end of the day. So, at this moment, the log file is compressed using the Mzip-Android\cite{mzipAndroid:2020:Online} library.

\begin{figure}
	\centering
	\includegraphics[width=0.25\textwidth]{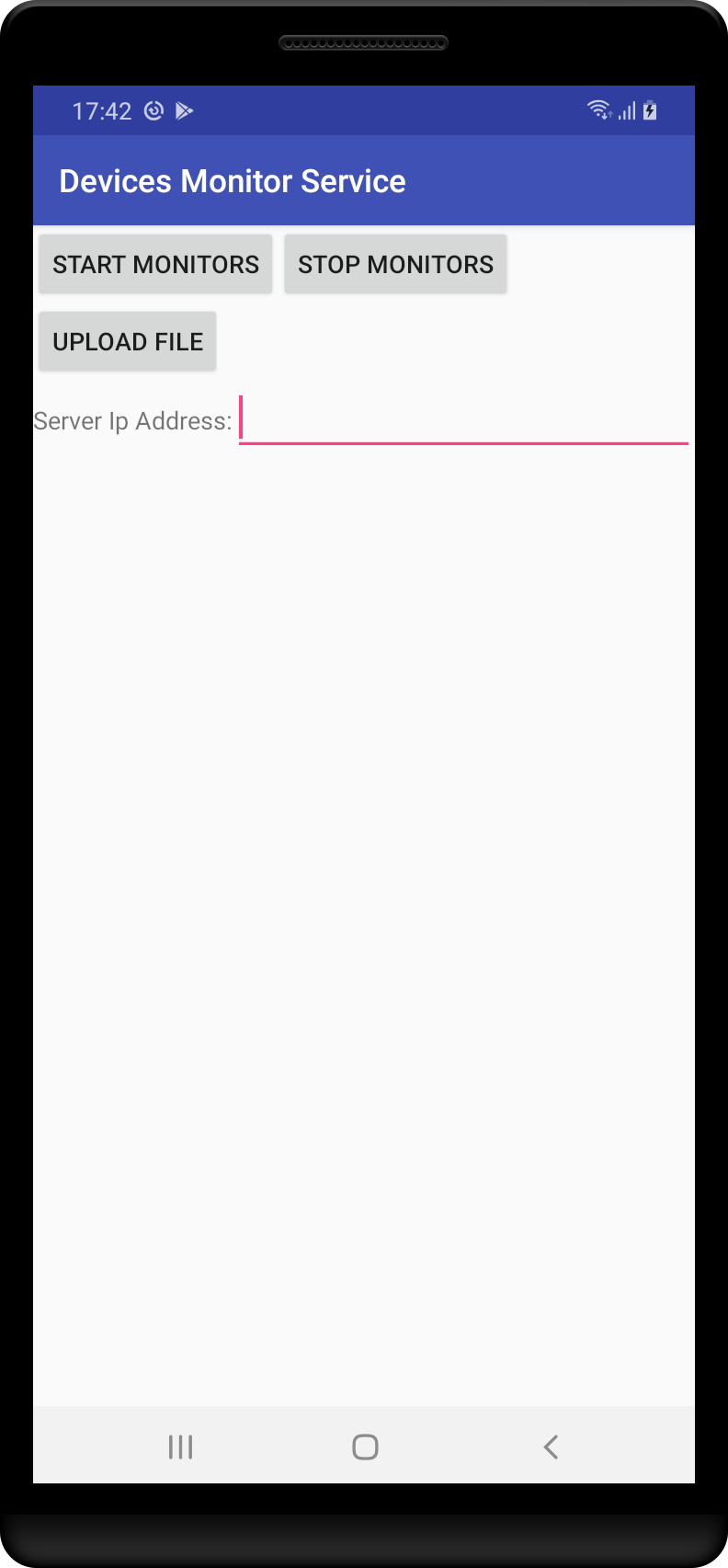}
	\caption{Monitor Service App}
	\label{fg:monitorService}
\end{figure}

After compressing the log file, the developed Android app sends the data to a pre-specified server that receives it and decompress it to analyze after. The layout of the developed Android app is shown in Figure \ref{fg:monitorService}.

The \textbf{Start Monitors} button is responsible for starting all the monitors inside the developed app. The \textbf{Stop Monitors} button stops all the monitors inside the developed app. The text field next to \textit{Server IP Address} label permits the user to set the address of the analysis server. Finally, the \textbf{Upload File} button is responsible for sending the previously stored user's usage log file using Android Upload Service\cite{androidUpload:2020:Online} and OkHTTP\cite{androidOkHttp:2020:Online} Android libraries.

To build the energy consumption models accurately, we need to measure the overhead in the smartphone's energy consumption due to the Devices Monitor Service app. To estimate the cost in the smartphone's energy consumption due to the developed app, we used the Energy Consumption profiler bash script listed in Listing \ref{lst:energyProfiler}.

To measure the Devices Monitor Service app overhead, we set the Samsung Galaxy A10 in airplane mode and performed 30 experiments with 120 samples in each state: without the app and with the app. The results fluctuated a lot, so we could not use the central limit theorem. 

According to Statdisk help site \cite{bootstrapHelp:2020:Online}, Bootstrap resampling is a procedure helpful in constructing confidence interval estimates of parameters when data sets have distributions that are far from Normal. So, we followed the method described in the Statdisk help site to calculate the confidence intervals of the smartphone's instant power means in each situation: with the Devices Monitor Service app running and without it. We constructed the confidence interval performing the bootstrap procedure with the 30 means obtained in the realized experiments. The obtained results are shown in Table \ref{table:appOverhadIC}.

\begin{table}[!ht]
    	\caption{App energy consumption overhead}
    	\centering
    	\begin{tabular}{ccc}
    		\hline
    		App State & CI Lower Bound& CI Upper Bound\\
    		\hline
    		Running & $561.4523mW$ & $665.759mW$\\
    		Stopped & $402.6635mW$ & $447.1311mW$ \\
    		\hline
    	\end{tabular}
        \label{table:appOverhadIC} 
\end{table}

From the results shown in Table \ref{table:appOverhadIC}, we can perceive that is some difference in smartphone's instant power when the Devices Monitor Service app is running, and we can estimate that it is the order of 200mW of overhead in instant smartphone power.

\subsection{Automatic Data Analysis Evaluation}
\label{sec:dataAnalysisEvaluation}

To implement the autonomous data analysis mechanism according to the procedure described in Section \ref{sec:automaticDataAnalysisMethod}, we have used the data structure and methods provided by the Pandas Data Analysis Library\cite{pandas:2019:Online}.

The primary motivation to use the Pandas Data Analysis Library is the fact that Python has been the most widely used programming language for performing data science tasks\cite{pythonDS:2020:Online}.

So, the analysis server receives the user's usage data, decompress it, and loads it using the Pandas library. For analyzing the user's usage data, we configured to load only a subset of the original variables collected, the ones that have numerical values and, besides that, the orientation one. The variables analyzed by the automatic data analysis mechanism are:

\begin{tasks}[label={\arabic*}](3)
\task CPU Frequency
\task CPU Usage
\task CPU Temperature
\task Mobile Signal\\Strength
\task Wi-Fi Signal\\Strength
\task Mobile RX Bytes
\task Mobile TX Bytes
\task Wi-Fi RX Bytes
\task Wi-Fi TX Bytes
\task Kb Read Per Second (Disk)
\task Kb Write Per Second (Disk)
\task Kb Read (Disk)
\task Kb Write (Disk)
\task Swap In
\task Swap Out
\task Context Switches
\task Red Mean
\task Red Std
\task Green Mean
\task Green Std
\task Blue Mean
\task Blue Std
\task Brightness
\task Orientation
\task Power(mW)
\end{tasks}

We have chosen only these variables because, during our tests, we have perceived that our regressors responsible for modeling the smartphone's energy consumption have a poor performance when we include many categorical variables.

With the filtered variables, including our target variable (Power in mW), we applied the steps of the Automatic model build methodology described in Section \ref{sec:automaticDataAnalysisMethod} to build the models that estimate the smartphone's energy consumption based on the user's usage pattern. The adopted procedures to implement the Automatic model building methodology and the results of the models' estimates are described in Section \ref{sec:automaticModelEvaluation}.

Using this procedure, we can automate the data analysis process and detect the variables that most impact the smartphone's energy consumption.

\subsection{Evaluation of energy Consumption model accuracy in biased data sets}
\label{sec:energyConumptionModelGeneralizationEvaluation}

To demonstrate that the methodology to construct an energy consumption model based on the user's usage can be used for any device, we follow the procedure stated in Section \ref{sec:energyConumptionModelGeneralizationMethod} utilizing the user's usage log generated by the Devices Monitor Service App.

To simulate the user's usage in a similar workload in different devices, we established two separate procedures to create 2 data sets: train and test data set.
The workload in each device is shown in Table \ref{table:trainWorkload}.

\begin{table}[!ht]
    	\caption{Simulated Main Workload-Train Data Set}
    	\centering
    	\begin{tabular}{ccc}
    		\hline
    		\multirow{2}{*}{App} & \multicolumn{2}{c}{Time(s)}\\
    		&Moto G6&Galaxy A10\\
    		\hline
    		Chrome & $8467$ & $634$\\
    		Youtube & $30884$ & $11929$ \\
    		\hline
    	\end{tabular}
        \label{table:trainWorkload} 
\end{table}

We realized an experiment with 47728 seconds of duration in Moto G6 and another experiment with 32063 seconds in Galaxy A10. We used many other apps during our experiments, but the most significant apps' usage is listed in Table \ref{table:trainWorkload}.

To establish a test workload, we realized another workload in each examined device with similar characteristics to the train one. The leading apps of the test workload are shown in Table \ref{table:testWorkload}.

\begin{table}[!ht]
    	\caption{Simulated Main Workload-Test Data Set}
    	\centering
    	\begin{tabular}{ccc}
    		\hline
    		\multirow{2}{*}{App} & \multicolumn{2}{c}{Time(s)}\\
    		&Moto G6&Galaxy A10\\
    		\hline
    		Youtube & $9350$ & $11625$ \\
    		\hline
    	\end{tabular}
        \label{table:testWorkload} 
\end{table}

During our experiments, we simulated a user with a usage pattern that demands more resources of the device. We watched two movies during train experiments in each device: "He's Just Not That into You" and "The Vow" and two films during test experiments in each device: "About Time" and "Vicky Cristina Barcelona."

Besides that, we have used other apps in a not coordinated way during the train and test experiments to provide the necessary subsidies to our chosen algorithms to build the smartphone's energy consumption model.

After collecting the data necessary to build the train and test data sets, we evaluated the data collected from the Moto G6 to identify the frequencies with which his Battery Fuel Gauge IC repeated the instant current values constructing a repetitions Numpy's array with them.

With the obtained Numpy's array, we go through the Power array of the Galaxy A10 smartphone train data set, replacing the actual values by the repeated ones in the same frequencies of the repetitions array. In this way, we have created a biased data set to test how one of our machine learning algorithms perform when the Battery Fuel Gauge IC is inaccurate.

To validate one of our machine learning algorithms, we have trained a Support Vector Regressor with the biased data set and measure its accuracy over the original test data set. To calculate the efficiency of the Support Vector Regressor Algorithm we have used the \textit{mean absolute error} metric of the scikit-learn\cite{scikit-learn} python library. The accuracy of the algorithm over the test data set was $289.678mW$.

To compare how well the obtained accuracy is, considering the temporal nature of the smartphone's instant power,  we have constructed a persistence regressor that predicts the next instant power repeating the last instant power from the test data set. So, in the persistence regressor $P[i+1]=P[i]$. At the end of the persistence regressor model construction, we have measured its accuracy over the test data set and obtained a \textit{mean absolute error} of $295.26mW$.

From these experiments, we have concluded that even in the presence of a biased data set, our machine learning regressor has better accuracy than the dummy one in $5,58mW$.

\section{Automatic model building evaluation}
\label{sec:automaticModelEvaluation}
The main contribution of this research is to establish a methodology to automatically build energy consumption models capable of estimate the smartphone's energy consumption based on the user's usage pattern. To aim this objective, as explained in Section \ref{sec:proposedMethod}, we needed to evaluate the relevant characteristics to define a consumption model based on the user's usage.

So, to evaluate the relevant characteristics to define a consumption model based on the user's usage pattern, we need to build an Android application and assess the models generated in a biased data set to establish the necessary subsidies to generalize our model construction strategy.

We choose to approach this problem of smartphone's energy consumption estimation based on the user's usage pattern through the construction of a methodology capable of creates energy consumption models automatically because these models are very dependent on the user's usage context.
In this section, we will discuss the strategies that we used to implement the Automatic model build methodology described in Section \ref{sec:automaticModelMethod} using the data collected to analyze the performance of the models in biased data set as shown in Tables \ref{table:trainWorkload} and \ref{table:testWorkload}.

\subsection{Data Loading and outliers removal}
\label{sec:dataLoading}

The first action performed by the automatic model building mechanism is data loading and outlier removal. It is an important step, according to the process defined by G\'{e}ron\cite{geron2019hands}, of the process because the user's usage data need to be prepared to be used by the smartphone's energy consumption estimators that will be constructed by the automatic model building mechanism.

\begin{figure}
	\centering
	\includegraphics[width=\textwidth]{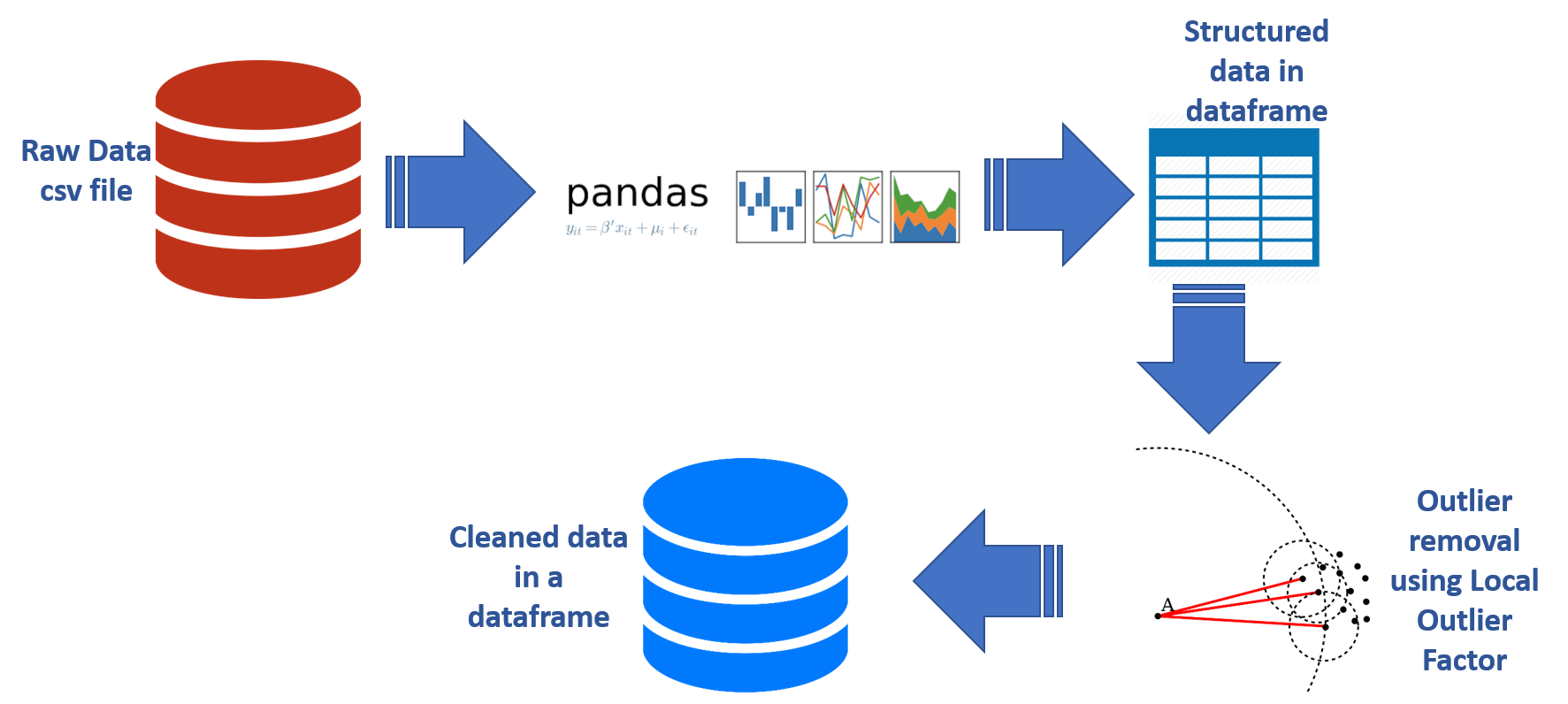}
	\caption{Data loading and outliers removal procedure}
	\label{fg:dataLoading}
\end{figure}

To implement this step, we followed the procedure depicted in Figure \ref{fg:dataLoading} and described bellow.

\MyPara{1.Load the raw CSV file to Pandas dataframe object:} In this step, we load the user's usage data using the \textbf{load\_csv} method of Pandas Data Analysis Library\cite{pandas:2019:Online}. After loading the CSV file, we make the transformations:
\begin{enumerate}
    \item Generate power in \textit{mW} multiplying the gathered current in \textit{ma} and the voltage in \textit{volts};
    \item Calculate the mobile signal strength using the relation $10^{gatheredSignalStrength/10}$;
    \item Select only the features that have \textbf{int64} and \textbf{float64} data types. 
\end{enumerate}

\MyPara{2.Outlier Removal using the Local Outlier Factor:} With the data in a structured way, we used the LocalOutlierFactor\cite{localOutlierFactor:2020:Online} module of Scikit-learn. We adopted the automatic scheme proposed by the module that permits it chooses the best algorithm to use.
\\
\MyPara{3.Cleaned data in a dataframe:} At the final of the data loading and outlier removal procedure, we returned the data in a cleaned way to apply the feature selection in the next step of our automated methodology to build the smartphone's energy consumption models.

\subsection{Feature Selection to train the user's usage model}
\label{sec:featuresSelection}
\begin{figure}
\centering
    \begin{subfigure}[t]{0.65\textwidth}
    \centering\includegraphics[width=\textwidth]{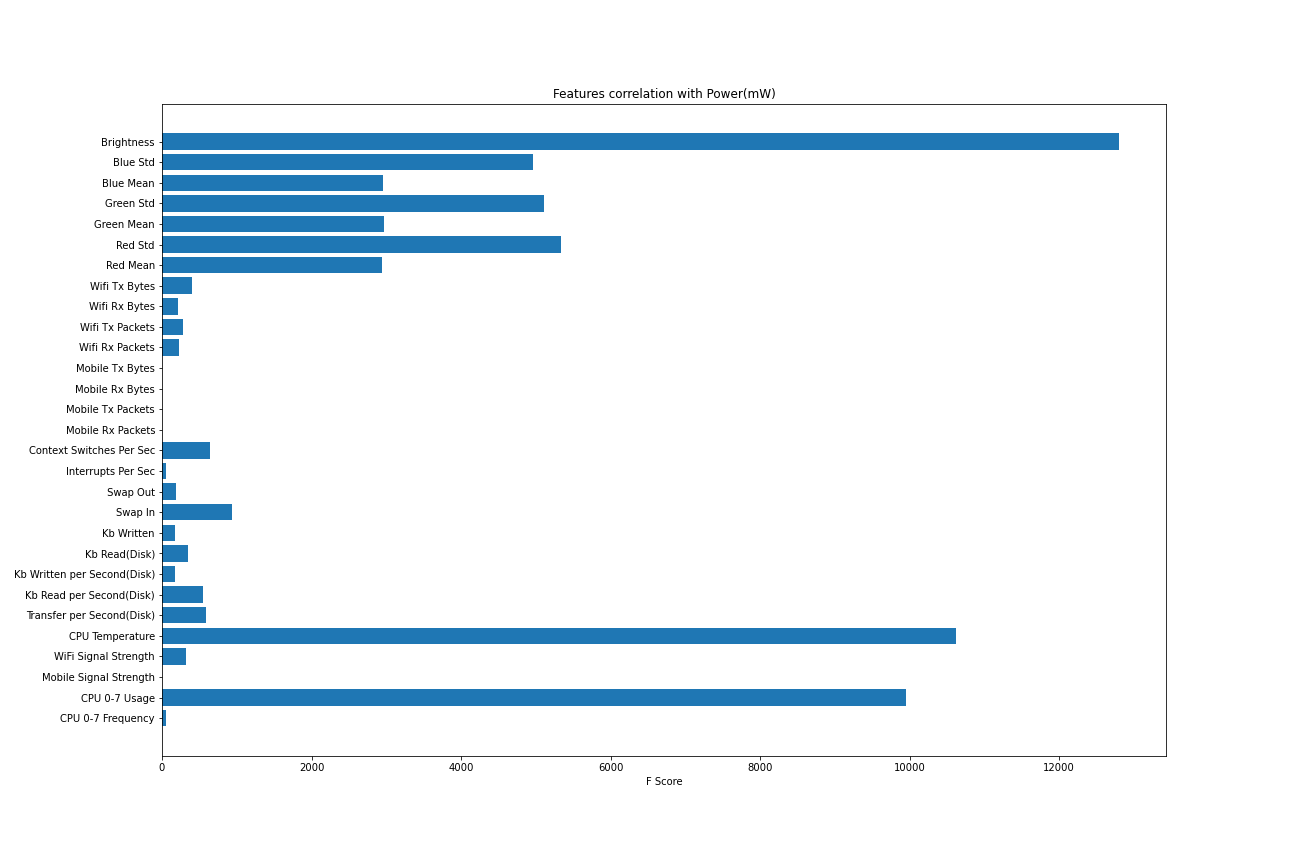}
    \caption{F-Test Strategy in Moto G6 Data}
    \label{fg:ftestMotoG6}
  \end{subfigure}
  \begin{subfigure}[t]{0.65\textwidth}
    \centering\includegraphics[width=\textwidth]{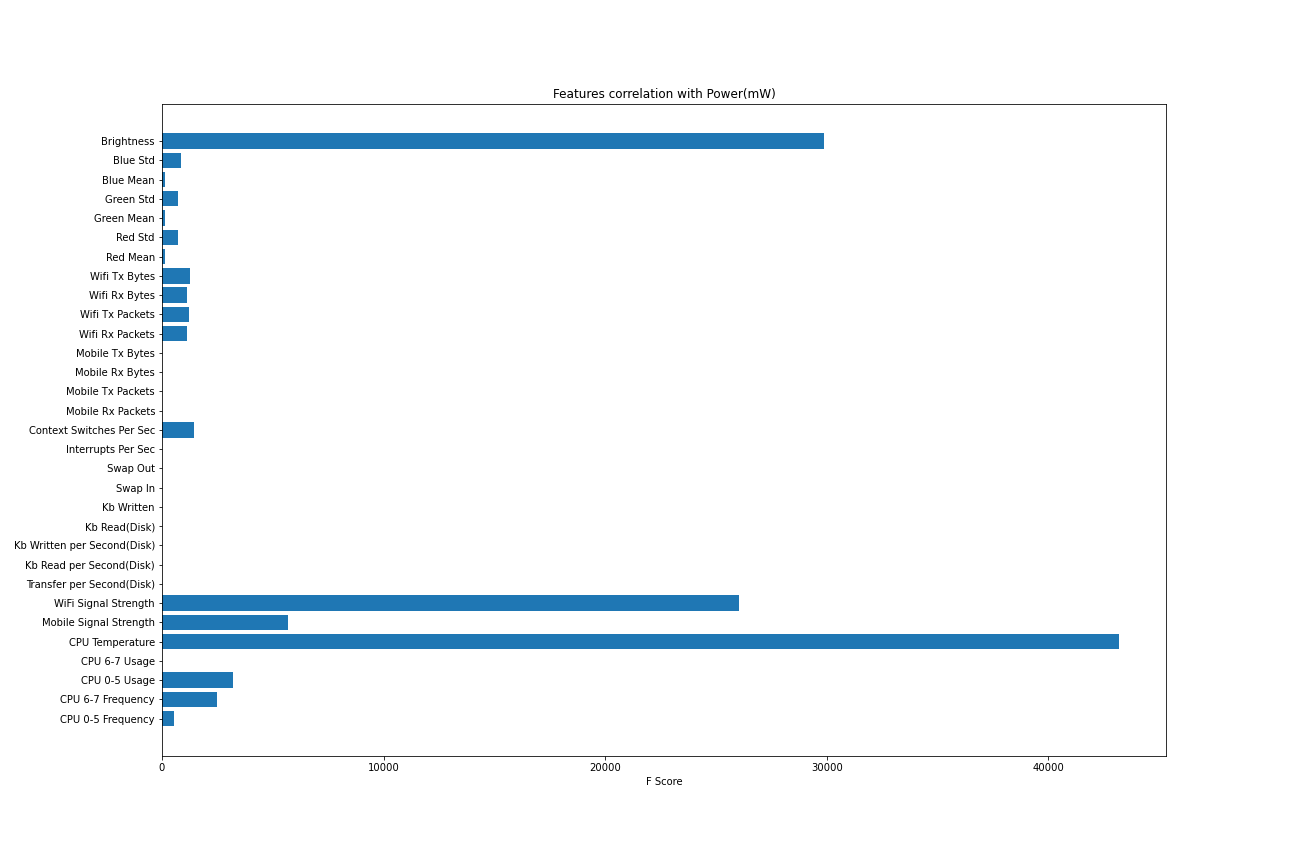}
    \caption{F-Test Strategy in Galaxy A10 Data}
    \label{fg:ftestGalaxyA10}
  \end{subfigure}
\end{figure}

In this step of our automatic model building methodology, we implemented the algorithm \ref{lst:bestKFeaturesAlgorithm} using the Python programming language. To implement the feature selection and chooses the best set of features to train the smartphone's energy consumption model based on the users' usage, we adopted the procedure below:

\MyPara{1.Choose the strategies to be evaluated:}To implement an automatic procedure to perform feature selection in the user's usage data, we decided to evaluate two score functions: F-Test and mutual information. During our experiments, we evaluated that, as mentioned before, all the categorical variables worse the model performance when we use them to select attributes. So, in our evaluation to build the automatic model construction methodology, we do not consider any of them. The results gathered when we select the F-Test\cite{fregression:2020:Online} strategy in Moto G6 and Galaxy A10 train data sets are shown in Figures \ref{fg:ftestMotoG6} and \ref{fg:ftestGalaxyA10}.

\begin{figure}
\centering
    \begin{subfigure}[t]{0.65\textwidth}
    \centering\includegraphics[width=\textwidth]{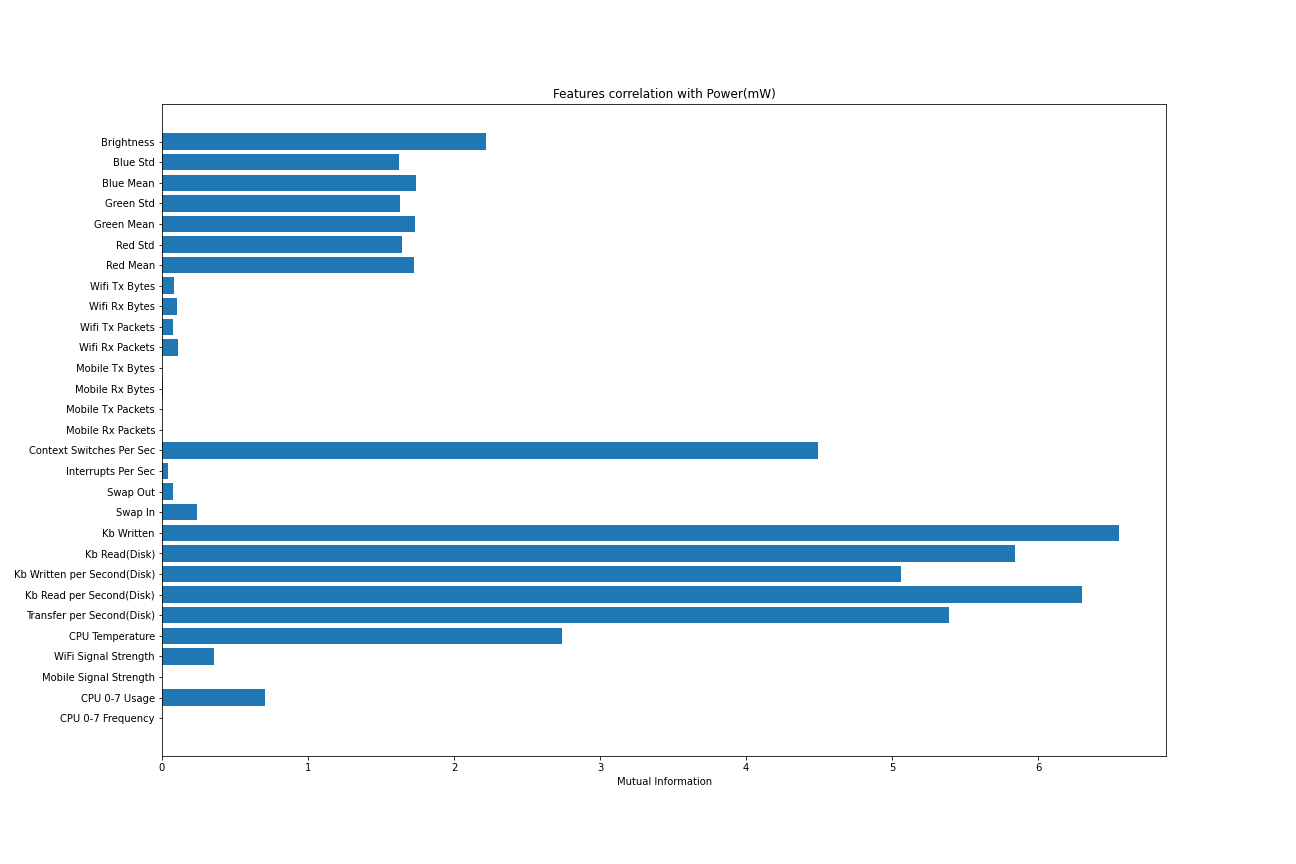}
    \caption{Mutual Information Strategy in Moto G6 Data}
    \label{fg:mutualInfoMotoG6}
  \end{subfigure}
  \begin{subfigure}[t]{0.65\textwidth}
    \centering\includegraphics[width=\textwidth]{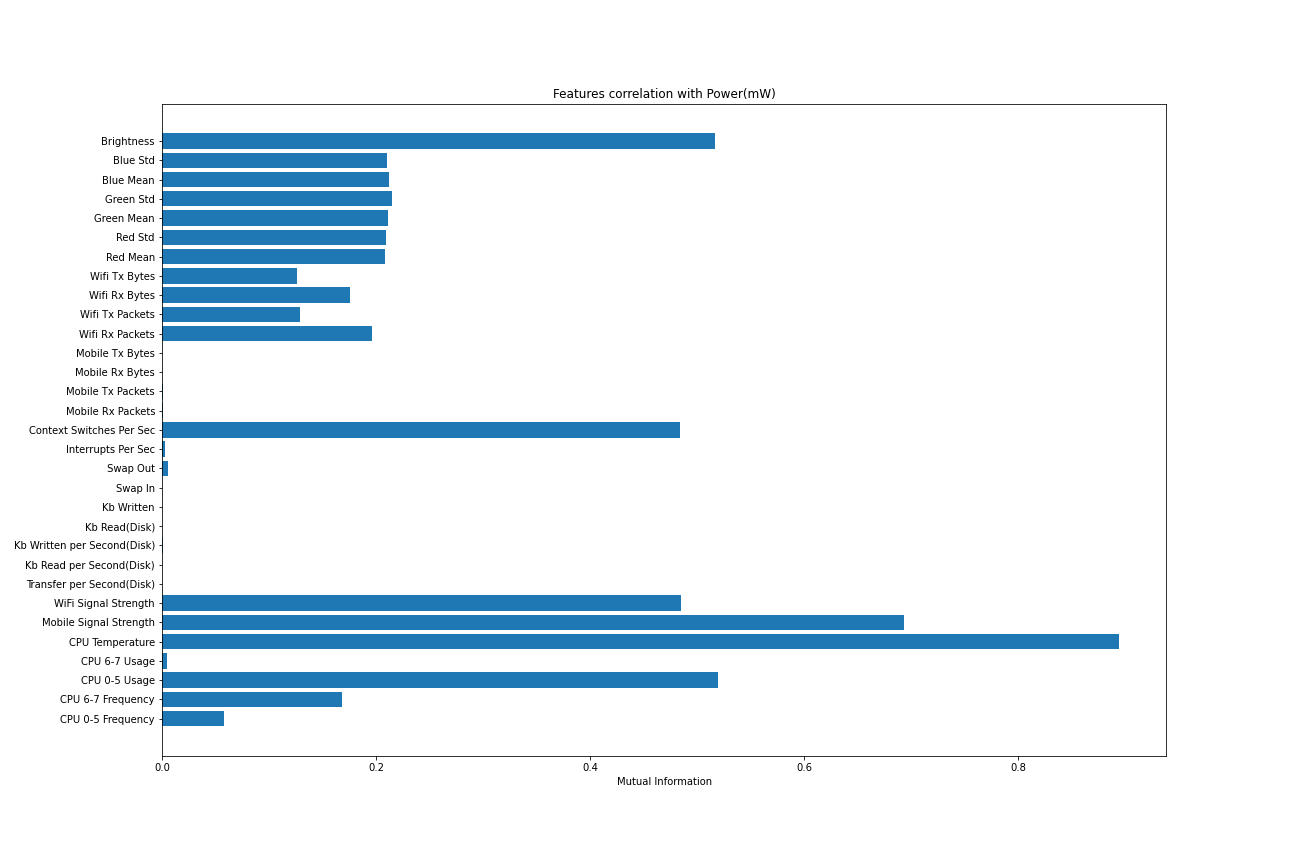}
    \caption{Mutual Information Strategy in Galaxy A10 Data}
    \label{fg:mutualInfoGalaxyA10}
  \end{subfigure}
\end{figure}

Another evaluated strategy was the Mutual Information because it does not assume a linear correlation among the features. Evaluating the Mutual Information strategy we gathered the results shown in Figures \ref{fg:mutualInfoMotoG6} and \ref{fg:mutualInfoGalaxyA10}.

At the end of this step, we verified that the best strategy to use was the mutual information because this one does not impose a linear correlation between the features (devices usage) and the target variable (power(mW)).

\MyPara{2.Choose the models that will be evaluated:} Another parameter that we have to evaluate to choose the best set of features to estimate the smartphone's user usage is the model that should be passed to test the various values of \textbf{k} best parameters to decide what is the best set of features.

In this step, we needed to evaluate the possibilities in a manual way to decide which algorithms will be used to select the best value of \textbf{k} because some of them consume much time to fit. Evaluating the possibilities, we choose two algorithms to be used in this step: \textbf{Random Forrest} and \textbf{Support Vector Regressor} because of their robustness and their time to fit.

\begin{figure}
\centering
    \begin{subfigure}[t]{0.65\textwidth}
    \centering\includegraphics[width=\textwidth]{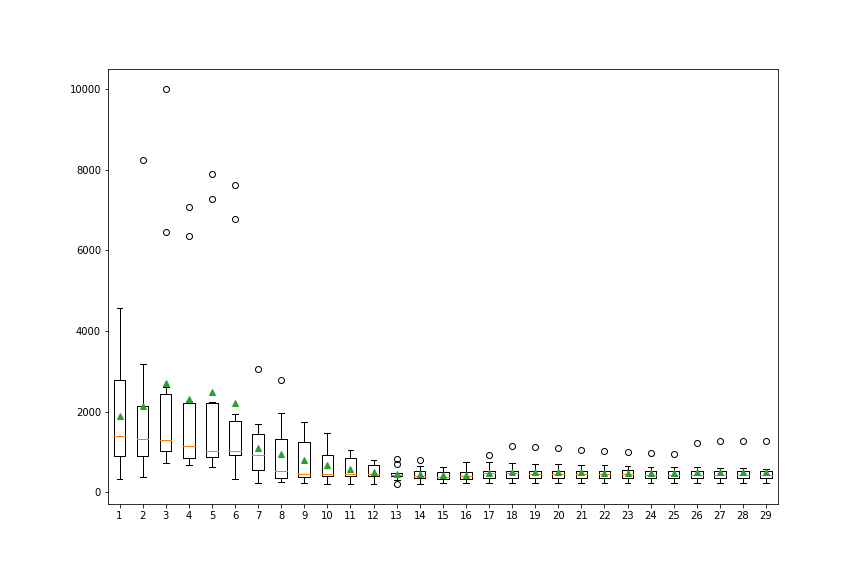}
    \caption{SVM model used to select features in Moto G6 Data}
    \label{fg:svmFselMotoG6}
  \end{subfigure}
  \begin{subfigure}[t]{0.65\textwidth}
    \centering\includegraphics[width=\textwidth]{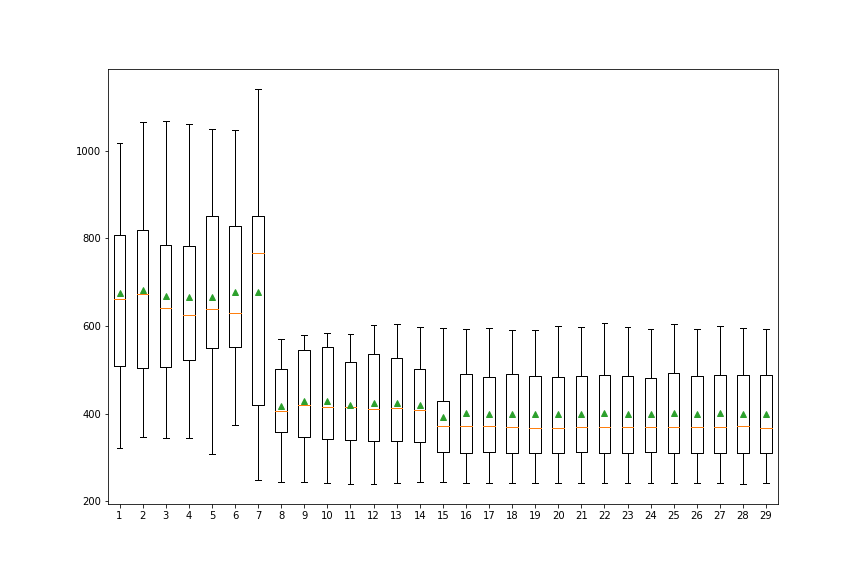}
    \caption{Random Forrest model used to select features in Moto G6 Data}
    \label{fg:rfFselMotoG6}
  \end{subfigure}
\end{figure}

The results gathered from applying these algorithms in the Moto G6  train data set are shown in Figures \ref{fg:svmFselMotoG6} and \ref{fg:rfFselMotoG6}.

\begin{figure}
\centering
    \begin{subfigure}[t]{0.65\textwidth}
    \centering\includegraphics[width=\textwidth]{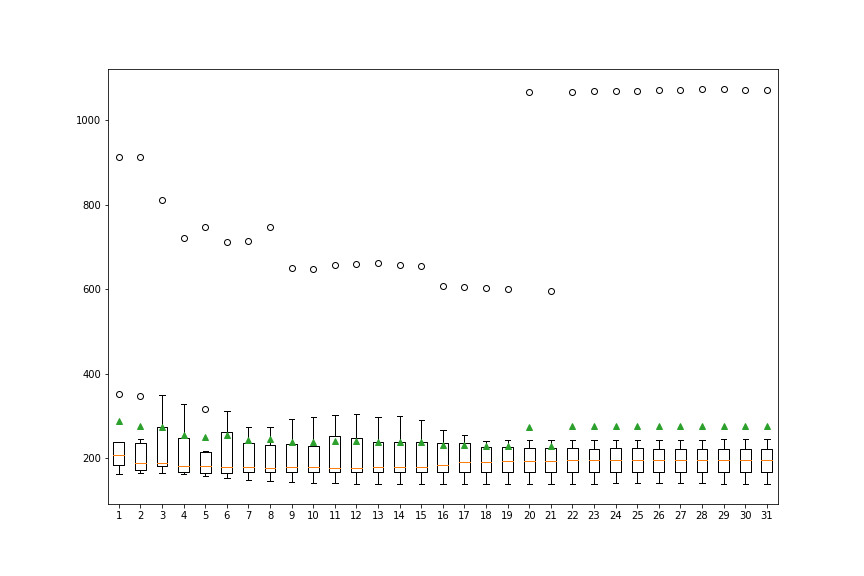}
    \caption{SVM model used to select features in Moto G6 Data}
    \label{fg:svmFselGalaxyA10}
  \end{subfigure}
  \begin{subfigure}[t]{0.65\textwidth}
    \centering\includegraphics[width=\textwidth]{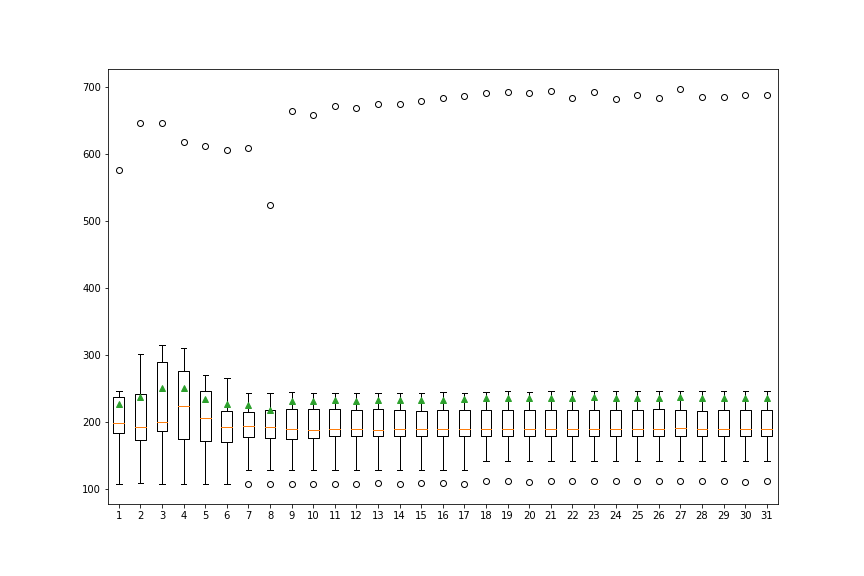}
    \caption{Random Forrest model used to select features in Moto G6 Data}
    \label{fg:rfFselGalaxyA10}
  \end{subfigure}
\end{figure}

Applying the selected algorithms in Galaxy A10 train data set we gathered the results shown in Figures \ref{fg:svmFselGalaxyA10} and \ref{fg:rfFselGalaxyA10}.

Applying these parameters to the algorithm \ref{lst:bestKFeaturesAlgorithm} we have found the values of \textbf{k} shown in Table \ref{table:kbest}

\begin{table}[!ht]
    	\caption{Best Set of features}
    	\centering
    	\begin{tabular}{ccc}
    		\hline
    		Data Set & Algorithm & Number of Features(K)\\
    		\hline
    		\multirow{2}{*}{Moto G6} & SVR &$18$\\
    		& Random Forrest &$18$\\
    		\hline
    		\multirow{2}{*}{Galaxy A10} & SVR &$18$\\
    		& Random Forrest &$19$\\
    		\hline
    	\end{tabular}
        \label{table:kbest} 
\end{table}

\subsection{Models Benchmark}
\label{sec:modelsBenchmark}

With the feature selection mechanism implemented in the previous step, we can implement the part of our autonomous model construction module responsible for benchmarking the various strategies used to build the energy consumption models.

In this stage, we implemented a procedure capable of evaluating many algorithms in three scenarios described in Table \ref{table:evaluatedScenarios}. We implemented the method in a flexible way that permits us to evaluate any model that we want.

\begin{table}[!ht]
    	\caption{Evaluated Scenarios}
    	\centering
    	\begin{tabularx}{\textwidth}{CC}    		
    	\hline
    		Scenario & Description\\
    		\hline
    		Full Features & In this scenario we considered all the numerical features and the orientation categorical one.\\
    		\hline
    		Without Orientation & In this scenario we considered only numerical features.\\
    		\hline
    		With Feature Selection & In this scenario we considered only numerical features and applied the feature selection procedure, as described before, to them.\\
    		\hline
    	\end{tabularx}
        \label{table:evaluatedScenarios} 
\end{table}
We evaluated all the selected algorithms using the method described in Section \ref{sec:modelsRepetitionMethod}. To perform the time series cross-validation we used the \textbf{TimeSeriesSplit}\cite{timeSeriesSplit:2020:Online} module of Scikit-learn library.

To calculate the median, the 25th percentile, and the 75th percentile, we used the percentile function of Numpy library\cite{van2011numpy}. 

After performing the procedures described in Section \ref{sec:modelsRepetitionMethod}, we serialize the each of the 30 chosen models in pickle files and the results gathered during this process in a JSON file for later use, in the step responsible for ranking the results.

\subsection{Evaluation of the generated energy consumption models}
\label{sec:energyConsumptionModelsEvaluation}

After building 30 models using each algorithm chosen, the automatic model building methodology can rank the models constructed. The ranking is performed using the scores that the models gathered when their estimations are compared with the real energy consumption in the test data set using the Nemenyi\cite{japkowicz2011evaluating} statistical test.

To aim this objective, we implemented the strategy to choose the most appropriate algorithm in each evaluated scenario, as shown in Table \ref{table:evaluatedScenarios}, following the procedure described below:

\MyPara{1.Load the results obtained in the previous step:} In this step, we iterate over the scenario directory containing the folders with the evaluated algorithms results in the respective scenario, and puts the results in a Python dictionary data structure. To test our methodology, we tested the algorithms listed bellow in each scenario listed in Table \ref{table:evaluatedScenarios}:

\MyPara{1.1.Full Features- Random Forrest:} To evaluate the Random Forrest algorithm with the full features, we adjusted its hyperparameters using the hpsklearn\cite{komer2014hyperopt} library. The hpsklearn tests this machine learning algorithm with all the hyperparameters using Bayesian Optimization.

\MyPara{1.2 Full Features-SVR:}To evaluate the Support Vector Regression algorithm with the full features, we adjusted its hyperparameters using the hpsklearn\cite{komer2014hyperopt} library. The hpsklearn tests this machine learning algorithm with all the hyperparameters using Bayesian Optimization.

\MyPara{1.3 Full Features-MLP:} To evaluate the Multilayer Perceptron neural network, we used the Tensorflow\cite{tensorflow2015-whitepaper} library to build a deep neural network architecture using dense and dropout layers. To optimize the hyperparameters of the Multilayer Perceptron networks constructed, we have used the Hyperas\cite{hyperas:2020:Online}. The Hyperas uses Bayesian hyperparameter optimization inside Keras using the Tree-structured Parzen Estimator\cite{bergstra2011algorithms} as the approach.

\MyPara{1.4 Full Features-LSTM:} To evaluate the LSTM neural network, we used the Tensorflow\cite{tensorflow2015-whitepaper} library to build a deep neural network architecture using lstm and dropout layers. To optimize the hyperparameters of the LSTM networks constructed, we have used the Hyperas\cite{hyperas:2020:Online} library.

\MyPara{1.5 Without Orientation:} In this scenario, we used the same algorithms and the same optimizations used in the full features scenario.

\MyPara{1.6 With Feature Selection- Random Forrest:} To evaluate the Random Forrest algorithm in this scenario, we used the default configuration of it in the Scikit-learn library. We decided to build this model in this way because we use the same model to select features and estimates smartphone energy consumption.

\MyPara{1.7. With Feature Section- SVR:} To evaluate the Support Vector Regression algorithm in this scenario, we used the default configuration of it in the Scikit-learn library. We decided to build this model in this way because we use the same model to select features and estimates smartphone energy consumption.

\MyPara{1.8. With Feature Selection-Neural Nets:} To evaluate a Neural Network in this scenario, we used the same features selected by the Support Vector Regression algorithm and the Hyperas\cite{hyperas:2020:Online} to choose one architecture among the options: LSTM, CNN, and MLP. Besides that, we used the Hyperas\cite{hyperas:2020:Online} library to optimize the hyperparameters of the chosen architecture.

\MyPara{1.9. With Feature Selection- LSTM:} To evaluate the LSTM neural network, we used the Tensorflow\cite{tensorflow2015-whitepaper} library to build a deep neural network architecture using lstm and dropout layers. To optimize the hyperparameters of the LSTM networks constructed, we have used the Hyperas\cite{hyperas:2020:Online} library.

\MyPara{2.Puts the scores of each iteration in a row:} In this step, we reorganize the loaded results in a matrix with the rows representing each of the 30 iterations and the columns representing each evaluated algorithm.

\MyPara{3.Rank data:} With the result matrix created in the previous step, we used the \textbf{rankdata}\cite{rankdata:2020:Online} method of the Scipy Stats python module.

\MyPara{4.Compute Critical Difference:} To evaluate if the algorithms rank difference has statistical significance, we need to calculate the Critical Distance. To calculate the Critical Distance, we have used the Orange Data Mining library to calculate the Critical Difference\cite{scoring:2020:Online}(CD) to get the most accurate algorithm in each data set for each studied smartphone. The CD for 30 experiments with a  $5\%$ significance level is $0.8563$.

\MyPara{5.Chooses the best scenario and algorithm to estimates the energy consumption:} Based on the results obtained during this procedure, the automatic methodology developed during this research can choose the best scenario and algorithm to estimates the smartphone's energy consumption based on the user's usage pattern.

So, using the procedures of the automatic model building methodology as described previously we coded the \textbf{Orientation} to be used in the \textbf{Full Features} scenario as \textbf{0} for \textbf{Portrait} and \textbf{1} for \textbf{Landscape}. The means of the results obtained from the 30 experiments carried out for each device following the procedures discussed in this Section and the previous ones are shown in Tables \ref{table:motoG6Results} and \ref{table:galaxyA10Results}.

\begin{table}[!ht]
    	\caption{Algorithms accuracy in Moto G6 Data Sets}
    	\centering
    	\begin{tabular}{ccc}
    		\hline
    		Scenario & Algorithm & Mean Absolute Error(mW)\\
    		\hline
    		\multirow{4}{*}{Full Features} & Random Forest &$339.6560$\\
    		& SVR &$341.8542$\\
    		& Deep MLP &$839.8552$\\
    		& LSTM &$158.5660$\\
    		\hline
    		\multirow{4}{*}{Without Orientation} & Random Forest &$355.1240$\\
    		& SVR &$895.9118$\\
    		& Deep MLP &$885.0611$\\
    		& LSTM &$188.9332$\\
    		\hline
    		\multirow{4}{*}{With Feature Selection} & Random Forrest &$429.907$\\
    		& SVR &$2684.955$\\
    		& Neural Nets &$1465.756$\\
    		& LSTM &$826.252$\\
    		\hline
    	\end{tabular}
        \label{table:motoG6Results} 
\end{table}

\begin{table}[!ht]
    	\caption{Algorithms accuracy in Galaxy A10 Data Sets}
    	\centering
    	\begin{tabular}{ccc}
    		\hline
    		Scenario & Algorithm & Mean Absolute Error(mW)\\
    		\hline
    		\multirow{4}{*}{Full Features} & SGD Regressor &$271.7107$\\
    		& SVR &$268.0919$\\
    		& Deep MLP &$261.4488$\\
    		& LSTM &$500.1657$\\
    		\hline
    		\multirow{4}{*}{Without Orientation} & Random Forest &$287.2655$\\
    		& SVR &$276.0355$\\
    		& Deep MLP &$443.6100$\\
    		& LSTM &$1387.6393$\\
    		\hline
    		\multirow{4}{*}{With Feature Selection} & Random Forrest &$299.125$\\
    		& SVR &$256.034$\\
    		& Neural Net &$623.056$\\
    		& LSTM &$356.352$\\
    		\hline
    	\end{tabular}
        \label{table:galaxyA10Results} 
\end{table}

In the results shown in Tables \ref{table:motoG6Results} and \ref{table:galaxyA10Results} the first machine learning algorithms for \textbf{Full Features} and \textbf{Without Orientation} scenarios are the ones chosen by the optimization process executed by the HyperOpt-Sklearn\cite{komer2014hyperopt} library and the \textbf{With Feature Selection} scenario presented in each table represents the features selected by the data analysis algorithm. We can perceive that in some situations, they have less accuracy than other algorithms because we have made the optimization over the train data sets and evaluated them over the test data sets. 

In Galaxy A10 data sets, we perceive the worse performance because, although we implemented mechanisms to avoid overfitting in the training data set, some algorithms overfitted. In Moto G6 data sets, we can perceive that, because of the bias that exists in the collected data, the feature selection algorithm does not be able to select the features with accuracy.

Based on the obtained results, we have computed the average ranks of them to perform the Nemenyi's\cite{japkowicz2011evaluating} statistical test to choose the best approach for each data set in each studied device. The results are shown in Tables \ref{table:motoG6Ranks} and \ref{table:galaxyA10Ranks}.

\begin{table}[!ht]
    	\caption{Algorithms average ranks in Moto G6 Data Sets}
    	\centering
    	\begin{tabular}{ccc}
    		\hline
    		Data Set & Algorithm & Average Rank\\
    		\hline
    		\multirow{4}{*}{Full Features} & Random Forest &$2$\\
    		& SVR &$3$\\
    		& Deep MLP &$4$\\
    		& LSTM &$1$\\
    		\hline
    		\multirow{4}{*}{Without Orientation} & Random Forest &$2$\\
    		& SVR &$3.5$\\
    		& Deep MLP &$3.5$\\
    		& LSTM &$1$\\
    		\hline
    		\multirow{4}{*}{With Feature Selection} & Random Forrest &$1.067$\\
    		& SVR &$3.97$\\
    		& Neural Nets &$2.84$\\
    		& LSTM &$2.133$\\
    		\hline
    	\end{tabular}
        \label{table:motoG6Ranks} 
\end{table}

\begin{table}[!ht]
    	\caption{Algorithms average ranks in Galaxy A10 Data Sets}
    	\centering
    	\begin{tabular}{ccc}
    		\hline
    		Data Set & Algorithm & Average Rank\\
    		\hline
    		\multirow{4}{*}{Full Features} & SGD Regressor &$2.7$\\
    		& SVR &$1.7$\\
    		& Deep MLP &$1.5$\\
    		& LSTM &$4$\\
    		\hline
    		\multirow{4}{*}{Without Orientation} & Random Forest &$2$\\
    		& SVR &$1$\\
    		& Deep MLP &$3.033$\\
    		& LSTM &$3.967$\\
    		\hline
    		\multirow{4}{*}{With Feature Selection} & Random Forrest &$2$\\
    		& SVR &$1$\\
    		& Neural Net &$4$\\
    		& LSTM &$3$\\
    		\hline
    	\end{tabular}
        \label{table:galaxyA10Ranks} 
\end{table}

Based on the average ranks obtained, we have used the Orange Data Mining library to calculate the Critical Difference\cite{scoring:2020:Online}(CD) to get the most accurate algorithm in each data set for each studied smartphone.

\begin{figure}
	\centering
	\includegraphics[width=0.9\textwidth]{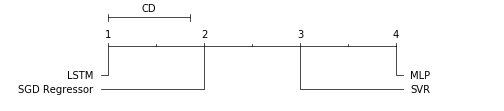}
	\caption{Moto G6- Full Features}
	\label{fg:motoG6FullFeatures}
\end{figure}

\begin{figure}
	\centering
	\includegraphics[width=0.9\textwidth]{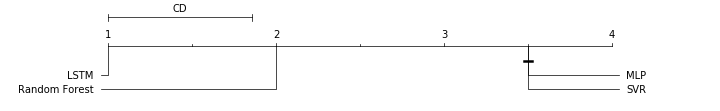}
	\caption{Moto G6- Without Observation}
	\label{fg:motoG6WO}
\end{figure}

\begin{figure}
	\centering
	\includegraphics[width=0.9\textwidth]{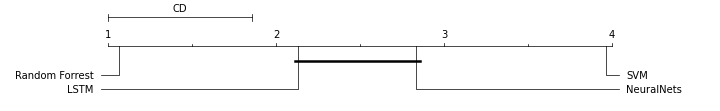}
	\caption{Moto G6- Best Features}
	\label{fg:motoG6Best}
\end{figure}

\begin{figure}
	\centering
	\includegraphics[width=0.9\textwidth]{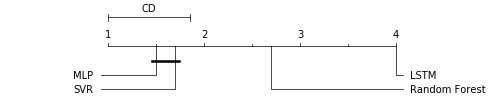}
	\caption{Galaxy A10- Full Features}
	\label{fg:galaxyA10FullFeatures}
\end{figure}

\begin{figure}
	\centering
	\includegraphics[width=0.9\textwidth]{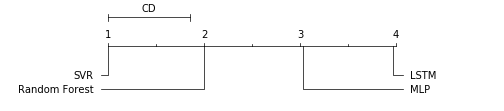}
	\caption{Galaxy A10- Without Observation}
	\label{fg:galaxyA10WO}
\end{figure}

\begin{figure}
	\centering
	\includegraphics[width=0.9\textwidth]{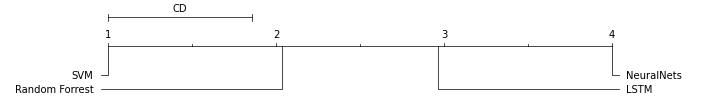}
	\caption{Galaxy A10- Best Features}
	\label{fg:galaxyA10Best}
\end{figure}

The CD for 30 experiments with a  $5\%$ significance level is $0.8563$. Using this critical distance and the obtained average ranks we can choose the most accurate algorithm in each data set for each studied smartphone automatically. The diagrams of the obtained results are shown in Figures \ref{fg:motoG6FullFeatures}, \ref{fg:motoG6WO},\ref{fg:motoG6Best},\ref{fg:galaxyA10FullFeatures}, \ref{fg:galaxyA10WO}, and \ref{fg:galaxyA10Best}.



Analyzing the Galaxy A10 graphs, we can perceive that for each situation, there is a most appropriate algorithm and that neither of them overlaps the others for all scenarios.

Observing the Moto G6 CD Diagrams, we can perceive that the LSTM networks overlap the other intelligent algorithms in two of the three evaluated scenarios. This fact shows that for biased Battery Fuel Gauge ICs, the LSTM presents an excellent performance. 

The LSTM architecture presents an excellent performance in biased data sets because, according to Sak \textit{et al.}\cite{sak2014long}, recurrent neural networks contain cycles that feed the network activations from a previous time step as inputs to the network to influence predictions at the current time step. These activations are stored in the internal states of the network, which can, in principle, hold long-term temporal contextual information. So, the LSTMs can remember the past situations and, in this case, estimates a consumption near the presented by the biased Battery Fuel Gauge IC.

\section{Research Questions and Discussion}
\label{sec:discussion}

\MyPara{RQ1:Is the user the primary workload for mobile systems?} The main goal of this research is to build a smartphone energy consumption model based on the user's usage pattern. So, before starting our research, we need to demonstrate which the study made by Shye \textit{et al.}\cite{shye2009into} is still valid. To test the user's usage pattern influence over the smartphone's energy consumption, we conducted two experiments using the same application with the same device doing different activities inside the app, as explained in Section \ref{sec:userInfluenceEnergyConsumptionMethod}. Performing these experiments, we can get some clues that Shye's study is still valid, and we can contribute to the research area of energy consumption modeling conducting research based on the user's usage pattern.
\\
\MyPara{RQ2:What the components with the most influence in the smartphone's energy consumption?} To develop an energy consumption modeling automated mechanism based on the user's usage pattern, we need to establish an automated tool to analyze the user's usage data and choose the most influential devices in smartphone energy consumption. So, we developed the algorithm explained in Section \ref{sec:dataAnalysisEvaluation} to accomplish this goal. The developed mechanism is capable of filter the interest data and apply the procedures to choose the variables which have the most influence in smartphone energy consumption. 
\\
\MyPara{RQ3:How well does our model generalize to any device?}
To develop an automated mechanism to modeling the smartphone energy consumption, we need to verify the accuracy of the developed modeling methodology over the biased data sets. This test is necessary because many devices gave an inaccurate Battery Fuel Gauge IC, and we propose a mechanism that uses this hardware as a basis to evaluate the smartphone's energy consumption and create a model. So, we have conducted an experiment where we biased an accurate data set, fitted our models using the biased data set, and evaluated their accuracy over an unbiased data set.
As we discussed in Section \ref{sec:energyConumptionModelGeneralizationEvaluation}, we demonstrated that our fitted models had overlapped the dummy regressor showing that our models can be fitted in any situation.

\MyPara{RQ4:How to assess the accuracy of the model built?} To create an autonomous energy consumption modeling mechanism, we need to define a methodology to evaluate the energy consumption models to choose the most appropriate model for each usage situation. According to the results shown in Tables \ref{table:galaxyA10Results} and \ref{table:galaxyA10Ranks}, we can perceive that neither model overlaps the other ones for all situations.

So, during our research, we developed a methodology that permits us to evaluate the most appropriate model automatically for predicting smartphone energy consumption in each situation.

Construct validity is threatened by the internal state of the Android operating system. We could perceive it during our experiments to determine the overhead of our developed Devices Monitor App. Although we have set the device in airplane mode during our tests, we have obtained results that fluctuated so much, indicating an overhead in some tests and not influencing the consumption in other ones.

So, the methods that suffer more influence of this fluctuation, such as the Neural Networks, tend to present many variations in their accuracy depending on the data set used to fit models.

These inaccuracies in devices' states and, consequently, in the device's energy consumption motivated us to evaluate our methodology to build energy consumption models in an inaccurate data set to demonstrate that our models can be used even in incorrect scenarios, as discussed in Section \ref{sec:energyConumptionModelGeneralizationEvaluation}.

Another limitation of the energy consumption models built from the user's usage pattern is the dependence of the user's behavior. Because of this dependence, we developed an automatic model building methodology that can retrain the smartphone's energy consumption models and adjust them according to the new situations.

\section{Conclusion and future works}
\label{sec:conclusion}

The development of smartphone technologies imposes many challenges for researchers. A prominent one is the optimization of energy consumption since the battery is a limited and scarce resource that does not evolve at the same pace as the smartphone's technology.

In this work, we developed a methodology to analyze the user behavior automatically and, based on it, build a smartphone energy consumption model that can be used by developers and autonomous optimizers. The main goal of this research is to establish a methodology to create an automatic mechanism to build energy consumption models for smartphones based on the user's usage that can be used by application developers and autonomous optimizers.

As future work, we intend to create automatic optimizers that consider the usage context to perform the optimizations based on it. The introduction of intelligence and situational awareness into the recognition process of human-centric event patterns could give a better understanding of human behaviors \cite{yurur2015generic}.








\end{document}